\documentclass[preprint]{elsarticle}
\usepackage{rotating}
\usepackage{graphicx}
\usepackage{amsmath,amssymb}
\usepackage[mathscr]{euscript}
\usepackage{color}
\usepackage{subfigure}
\begin{document}
\title{Deviations from Born-Oppenheimer mass scaling \\
in spectroscopy and ultracold molecular physics}
\author[jqc]{Jesse J. Lutz\fnref{fn1}}
\ead{jesse.lutz.ctr@afit.edu}
\author[jqc]{Jeremy M. Hutson\corref{cor1}}
\ead{j.m.hutson@durham.ac.uk}
\fntext[fn1]{Present address: Department of Engineering Physics, Air Force
Institute of Technology, 2950 Hobson Way, Wright-Patterson AFB, OH 45433-7765, USA}
\cortext[cor1]{Corresponding author}
\address[jqc]{Joint Quantum Centre (JQC) Durham-Newcastle, Department of
Chemistry, Durham University, South Road, Durham DH1 3LE, United Kingdom}

\begin{abstract}
We investigate Born-Oppenheimer breakdown (BOB) effects (beyond the usual mass
scaling) for the electronic ground states of a series of homonuclear and
heteronuclear alkali-metal diatoms, together with the Sr$_2$ and Yb$_2$
diatomics. Several widely available electronic structure software packages are
used to calculate the leading contributions to the total isotope shift for
commonly occurring isotopologs of each species. Computed quantities include
diagonal Born-Oppenheimer corrections (mass shifts) and isotopic field shifts.
Mass shifts dominate for light nuclei up to and including K, but field shifts
contribute significantly for Rb and Sr and are dominant for Yb. We compare the
{\em ab initio} mass-shift functions for Li$_2$, LiK and LiRb with
spectroscopically derived ground-state BOB functions from the literature. We
find good agreement in the values of the functions for LiK and LiRb at their
equilibrium geometries, but significant disagreement with the shapes of the
functions for all 3 systems. The differences may be due to contributions of
nonadiabatic terms to the empirical BOB functions. We present a semiclassical
model for the effect of BOB corrections on the binding energies of
near-threshold states and the positions of zero-energy Feshbach resonances.
\end{abstract}

\date{\today}
\maketitle

\section{Introduction}
\label{sec:intro}

The Born-Oppenheimer approximation (BOA) lies at the heart of chemical and
molecular physics. It underpins the concepts of potential energy curves and
surfaces that are universally used to understand and interpret molecular
structure and dynamics. For many purposes, the BOA is adequate and it is not
necessary to go beyond it. However, for light nuclei and high-precision work,
deviations from the BOA are important. Quantitative investigations of such
deviations go back at least to the theoretical work of Ko{\l}os and Wolniewicz
on H$_2$ \cite{Kolos:1964, Kolos:1965}, which stimulated reinterpretation of
the experimental spectrum by Herzberg \cite{Herzberg:1970}. They are also
important for H$_3^+$ \cite{Polyansky:1999}, and they have been characterized
spectroscopically for hydrides such as HeH$^+$ \cite{Coxon:HeH+:1999}, BeH$^+$
\cite{Coxon:BeH+:1997}, HF \cite{LeRoy:1999, Coxon:HF:2006, Coxon:HX:2015}, HCl
\cite{Coxon:HCl:2000, Coxon:HX:2015}, HBr and HI \cite{Coxon:HX:2015}, AgH
\cite{LeRoy:2002, LeRoy:2005}, LiH \cite{Coxon:LiH:2004}, BeH \cite{LeRoy:2006}
and MgH \cite{Henderson:2013} and for CO \cite{Coxon:CO:2004}, Li$_2$
\cite{WMLjcp2002, Coxon:2006, LeRoy:2009}, LiK \cite{Tiemann:2009}, and LiRb
\cite{Ivanova:2011}. For molecules without such light nuclei, the deviations
have been hard to detect \cite{Seto:2000, Docenko:2007, Strauss:2010}, although
indications of them have been seen in K$_2$ \cite{Falke:2008}, Rb$_2$
\cite{Seto:2000,VKKpra2009}, and I$_2$ \cite{Knockel:2004,Salumbides:2008}.

Recent developments in the field of ultracold atoms and molecules offer a new
stimulus to understand deviations from the BOA. Key quantities in this field
are the binding energies of levels very close to dissociation and the positions
of zero-energy Feshbach resonances as a function of magnetic field
\cite{Chin:RMP:2010}. The latter are essentially the fields at which the
energies of bound molecular states exactly equal those of free atoms. For pairs
of heavy atoms, potential curves derived from one isotopolog have been very
successfully used to predict resonance positions for another by simply
rerunning the scattering calculations with a different reduced mass (and
different atomic properties such as nuclear spins and hyperfine splittings)
\cite{Simoni:2008, Cho:RbCs:2013, Blackley:85Rb:2013}. However, Julienne and
Hutson \cite{Julienne:Li67:2014} have recently shown that deviations from the
BOA are responsible for 4~G of the shift in resonance position between $^6$Li+$^6$Li
and $^7$Li+$^7$Li, and have obtained potential curves that include the necessary
corrections for both the singlet and triplet states.

Breakdown of the BOA is also crucial for attempts to use the spectroscopy of
ultracold molecules to explore fundamental physics. For example, Kitagawa {\em
et al.}\ \cite{Kitagawa:2008} have measured the binding energies of
near-dissociation states of several isotopologs of Yb$_2$ and similar
experiments are underway for Sr$_2$
\cite{Zelevinsky:2006,Stellmer:2012,McGuyer:2013,Borkowski:2014,McGuyer:2014}.
For Yb$_2$, binding energies are generally in good agreement with the
predictions of Born-Oppenheimer mass scaling. However, there are proposals to
use the small deviations from such scaling to place limits on the magnitude of
a ``fifth force" that may exist in addition to the familiar electromagnetic,
gravitational and strong and weak nuclear forces (see, e.g., Refs.
\cite{Salumbides:2013,Niu:2014,Salumbides:2014}). Before any such effects can
be ascribed to novel physics, it is crucial first to consider deviations from
mass scaling that arise from conventional physics.

The purpose of the present paper is to explore the capabilities of current
theoretical methods for calculating deviations from the BOA, and to investigate
their magnitude for species of importance in the study of ultracold molecules.
These include both homonuclear and heteronuclear alkali-metal diatomics and
molecules such as Sr$_2$ and Yb$_2$. For the heavier molecules, almost nothing
is known about the corrections needed, and even order-of-magnitude estimates
are valuable. We consider effects due to both finite nuclear mass
(isotopic mass shifts) and finite nuclear volume (isotopic field shifts).

The structure of this paper is as follows. Section \ref{sec:theo} describes the
theory underlying isotope shifts and outlines previous work, and section
\ref{sec:RandD} describes the software packages and approximations used in our
work and presents the results and discussion.

\section{Theory}
\label{sec:theo}

Electronic structure theory provides a framework for computing isotope shifts.
Atomic and molecular calculations are usually performed assuming {\it a priori}
that the nuclei are both infinitely heavy and infinitely tiny, i.e., they are
treated as point charges. The former approximation is commonly called the BOA
although it resembles the treatment of Born and Huang \cite{BHbook1956}
more than it does the original proposition of Born and Oppenheimer \cite{BOap1927}.

For atomic systems, the theory of isotopic shifts has been rigorously developed
within a relativistic formalism \cite{GNRjpb2011} and dedicated programs are
widely available for their {\em ab initio} calculation (see, e.g., Refs.\
\cite{NGGGJcpc2013,JGBcpc2013}). Meanwhile in molecular systems,
bonding-induced isotopic shifts produce relatively small effects in
spectroscopic results. To help diagnose whether isotopic mass shifts are
important in a given application, Born and Huang derived a first-order
perturbative correction to the BOA energy \cite{BHbook1956}, which can take the
form,
\begin{equation}
\Delta V^{\rm ad}_{\alpha,k}(\boldsymbol{R}_{\mathrm n})
=\sum_A \langle\Psi_k(\boldsymbol{r};\boldsymbol{R}_{\mathrm n})|\hat{T}_A|
\Psi_k(\boldsymbol{r};\boldsymbol{R}_{\mathrm n})\rangle,
\label{BH_DBOC}
\end{equation}
where $\alpha$ indicates the isotopolog,
$\hat{T}_A=-\frac{\hbar^2}{2M_A}\hat{\nabla}^2_A$ is the kinetic energy
operator for nucleus $A$ and $\Psi_k(\boldsymbol{r};\boldsymbol{R}_{\mathrm
n})$ is the normalized electronic wavefunction for state $k$ obtained within
the BOA, with explicit dependence on electronic positions $\boldsymbol{r}$ and
parametric dependence on nuclear positions $\boldsymbol{R}_{\mathrm n}$. The
quantities $\Delta V_{\alpha,k}^{\rm ad}(\boldsymbol{R}_{\mathrm n})$ are known
as adiabatic corrections or diagonal Born-Oppenheimer corrections (DBOCs).

The approach embodied by Eq.\ \ref{BH_DBOC} appears to be problematic from a
formal perspective, because the equations involve manipulating continuum
functions as if they were normalizable. Fortunately, as was later shown by
Kutzelnigg \cite{Kutzelnigg1997}, this form of the adiabatic correction is
correct and, in fact, can be derived rigorously by avoiding the actual
specification of relative coordinates in the center-of-mass separation.

The equations of the pragmatic ansatz of Eq.\ \ref{BH_DBOC} were first solved
using Hartree-Fock (HF) wavefunctions by Sellers and Pulay \cite{SPcpl1984},
and later by Handy {\em et al.}\ \cite{Handy1986a}. The expressions are
evaluated by holding all but one nucleus fixed and calculating analytic
derivatives of the wavefunction with respect to the coordinates of the
remaining nuclei. The significance of the correction was established by early
investigations of its effect on molecular bond lengths and vibrational
frequencies \cite{Sellers,Handy1996a}, thermochemical reaction barrier heights
\cite{Truhlar,Handy1996b} and the singlet-triplet gap in methylene
\cite{Handy1986b}. Adiabatic corrections for water were a critical component of
models that demonstrated the existence of water on the sun
\cite{BWjcp1978,water1,water2,water3}. Computing adiabatic corrections at the
HF level has by now become so routine that they are included in standard
composite methods for high-accuracy thermochemical calculations
\cite{HEAT1,HEAT2,W4,W2F12}.

Valeev and Sherrill showed that the inclusion of electron correlation via
configuration interaction in the wavefunction can lead to changes in the
absolute DBOC of a few percent for some systems, with the most pronounced
effects occurring in hydrides \cite{VSjcp2002,TVSjpca2004}. When changes with
respect to geometry are considered instead, correlation effects can contribute
much more significantly \cite{GLRcpl2001}, in the same way as correlation can
contribute more significantly to relative than to total energies. The CFour
package \cite{CFour} has made available codes for the analytic evaluation of
adiabatic corrections using wavefunctions from coupled-cluster (CC)
calculations \cite{GTKjcp2006} and M{\" o}ller-Plesset perturbation theory
\cite{MPDBOC}.
Schwenke also evaluated adiabatic and non-adiabatic corrections
\cite{Schwenke1} using internally contracted multireference CI wavefunctions
\cite{Schwenke2,Schwenke3}, but unfortunately his program was not widely
distributed. It is reported \cite{IASHjpca2016} that the next release of the
GAMESS software package will have the capability to compute DBOCs based on
scalar-relativistic Hamiltonians.

Nonadiabatic corrections, originating from the off-diagonal matrix elements of
the nuclear kinetic energy operator, can be essential for understanding
molecular dynamics when different electronic states come close together
\cite{Tbook1976}. Even for nondegenerate electronic states, they can make
significant contributions to spectroscopic line positions, as discussed below.
Nevertheless, an adiabatic representation is convenient because it retains the
concept of a potential energy surface \cite{IUPACCCT,Bbbo2006} and can be
obtained using standard analytic derivative techniques
\cite{Handy1986a,GTKjcp2006}. For these reasons, we focus here on obtaining
adiabatic corrections, though we note that good progress has recently made for
obtaining highly accurate nonadiabatic corrections by Pachucki and Komasa
\cite{PKjcp2008,PKjcp2009}.

There is a fundamental difference between the interpretations of the
Born-Oppenheimer approximation in common use among spectroscopists and
electronic structure theorists. These are most easily illustrated by
considering the two different treatments of a diatomic molecule.

Electronic structure theorists normally consider {\em nuclei} moving on the
potential energy curves or surfaces. For a diatomic molecule, the internuclear
distance is $R_{\rm n}$ and the reduced mass for nuclear motion is $\mu_{\rm
n}=M_1M_2/(M_1+M_2)$, where $M_1$ and $M_2$ are nuclear masses. This separation
gives the form of the Born-Oppenheimer approximation described above. However,
it has the disadvantage that the nonadiabatic corrections are nonzero even as
$R_{\rm n}\rightarrow\infty$, and this presents problems for scattering theory
because the asymptotic wavefunctions are not simple products of the wave
functions of the separated atoms.

In spectroscopy and scattering theory, by contrast, it is common to consider
{\em atoms} moving on effective potential energy curves. Electrons are
considered to be parts of the atoms and to move with them. This makes good
physical sense, at least for core electrons, which are tightly bound to the
nuclei. For a diatomic molecule, the reduced mass for atomic motion is
$\mu_{\rm a}=M_{\rm 1a}M_{\rm 2a}/(M_{\rm 1a}+M_{\rm 2a})$, where $M_{\rm 1a}$
and $M_{\rm 2a}$ are atomic masses.

The formal justification of this approach is based on work by Bunker and Moss
\cite{Bunker:effective:1977} and Watson \cite{Watson:1980}. Bunker and Moss
derived an effective Hamiltonian for a single electronic state of diatomic
molecule, taking nonadiabatic couplings into account by means of a contact
transformation. Their treatment introduces nonadiabatic corrections through
both an effective potential term $\Delta W(R)$ (which scales as $\mu^{-2}$) and
separate $R$-dependent reduced masses $\mu_{\rm vib}(R)$ and $\mu_{\rm rot}(R)$
for the vibrational and rotational motion, which differ from $\mu_{\rm n}$ by
terms that scale as $\mu^{-1}$. In subsequent work, Bunker, McLarnon and Moss
\cite{Bunker:H2:1977} showed that, if the $R$-dependence is neglected, the
value of $\mu_{\rm v}$ that gives an optimum fit to the full nonadiabatic
energies of H$_2$ is closer to $\mu_{\rm a}$ than to $\mu_{\rm n}$. Watson
\cite{Watson:1980} showed that the effective Hamiltonian of Ref.\
\cite{Bunker:effective:1977} may be rearranged to a form containing the
``charge-modified reduced mass" $\mu_{\rm C}=M_{\rm 1a}M_{\rm 2a}/(M_{\rm
1a}+M_{\rm 2a}-m_eQ)$, where $Q$ is the molecular charge; $\mu_{\rm C}$ reduces
to $\mu_{\rm a}$ for a neutral molecule. The replacement of $\mu_{\rm n}$ with
$\mu_{\rm a}$ substantially reduces the remaining nonadiabatic corrections if
the electrons move with their respective nuclei; that is, $\mu_{\rm vib}(R)$
and $\mu_{\rm rot}(R)$ are much closer to $\mu_{\rm a}$ than to $\mu_{\rm n}$.
It essentially corresponds to a change of viewpoint: if $\mu_{\rm a}$ is used
as the reduced mass, then the nonadiabatic terms account for the extent to
which the electrons fail to move with their ``parent" atom, whereas if
$\mu_{\rm n}$ is used then the nonadiabatic terms need to account for the
extent to which the electrons move with the nuclei at all.

Watson \cite{Watson:1980} showed that it is not possible to determine both the
adiabatic correction and the nonadiabatic correction terms in $\mu_{\rm vib}$
simultaneously from transition frequencies alone, and gave an expression for an
effective adiabatic correction that absorbs the nonadiabatic corrections in
$\mu_{\rm vib}$ (but not those in $\mu_{\rm rot}$).

Watson's nonadiabatic corrections become asymptotically zero when the electrons
of each atom move with it and the atomic reduced mass is used. However, his
effective adiabatic corrections are {\em not} necessarily asymptotically zero.
Nevertheless, we are free to choose the zero of energy for each isotopolog,
and it is convenient to choose it as the energy of the free atoms at the
threshold of greatest interest (which is the ground-state dissociation energy
in the present work). With this choice, the value of $\Delta V_{\alpha,0}^{\rm
ad}$ at infinity for each isotopolog is absorbed into the definition of the
origin, and only $\Delta V_{\alpha,0}^{\rm ad}(R)-\Delta V_{\alpha,0}^{\rm
ad}(\infty)$ is explicitly included as an adiabatic correction. Pachucki and
Komasa \cite{PKjcp2008,PKjcp2009} give a perturbative treatment of nonadiabatic
effects that reaches the same conclusion. Non-zero adiabatic corrections at
infinity are still required for any electronic states that dissociate to
different limits.

In the context of electronic structure theory, Handy and Lee \cite{Handy1996a}
recommended that atomic masses be used rather than nuclear masses when
computing adiabatic corrections. Kutzelnigg \cite{Kutzelnigg2007} also
discussed this question in detail, and concluded that at least inner-shell
electrons should be considered to move with the nuclei. Nevertheless, the usual
convention in electronic structure theory is to use nuclear masses
\cite{GLRcpl2001}, and we follow that convention in the remainder of this
paper.

For atomic systems, the quantity that corresponds to the molecular adiabatic
correction is the nuclear mass shift. Traditionally, the total atomic mass
shift was separated into a {\em normal mass shift} and a {\em specific mass
shift} \cite{Hughes:1930}. The normal mass shift is obtained simply by
replacing all electron masses with reduced masses $\mu_e=m_e m_{\rm n}/(m_e +
m_{\rm n})$ in the calculation, and results in a simple scaling of all atomic
state energies (and transition frequencies) by a factor $\mu_e/m_e$. The
specific mass shift, however, varies from state to state and its calculation
involves mass polarization terms written in terms of products of momentum
operators on pairs of electrons. More recently, it has been recognized that
this approach can be unreliable, and that a more complete result may be
obtained in terms of the relativistic nuclear recoil operator
\cite{Fricke:1973}. At least in the nonrelativistic case, the adiabatic
correction as evaluated by electronic structure programs such as CFour is
asymptotically equivalent to the sum of the total mass shifts of the
constituent atoms.

For heavy atoms, isotope shifts are usually dominated by the nuclear field
shift, which results from the finite volume of the nucleus, rather than by the
mass shift. In quantum mechanics, electrons can penetrate nuclei, and the
electric potential they experience inside the nucleus is less negative than
$-Ze/(4\pi\epsilon_{0}r)$. Consequently, the energies of penetrating orbitals
are shifted upwards due to the finite size of the nucleus. The magnitude of
this effect depends on the structure of the nucleus involved, but can be shown
to a rough approximation \cite{OttenBook} to scale with the atomic number $Z$
and mass number $A$ as
\begin{equation}
\delta E^{AA^{\prime}}_{\mathrm FS} \propto \frac{Z^2}{\sqrt[3]{A}}.
\end{equation}
In atomic spectroscopy the interplay between mass and field shifts has been
studied extensively and a crossover point is estimated to occur at $Z\approx38$
\cite{BDNps2013}.

The theory of isotopic field shifts was first formulated for atomic systems by
Rosenthal and Breit \cite{RBpr1932} and by Racah \cite{Rn1932} in 1932.
Assuming a spherical nuclear charge distribution and performing a power
expansion of the electron density within the nucleus,
the conventional model gives the first-order perturbative correction to the
energy shift of level $i$ in going from isotope $A$ to $A^{\prime}$ for an atom
as
\begin{equation}
\delta E_i^{(1)A,A^{\prime}}
= \frac{2\pi}{3}Z\left(\frac{e^2}{4\pi\epsilon_0}\right)
  |\Psi(0)|^2_i \lambda^{A,A^{\prime}}.
\label{IFS1}
\end{equation}
Here $|\Psi(0)|^2_i$ is the density at the nucleus, often known as the contact
density, and to a first approximation the so-called ``nuclear parameter'',
$\lambda^{A,A'}$, is
\begin{equation}
\lambda^{A,A^{\prime}} = \delta\langle r^2 \rangle^{A,A^{\prime}}
= \langle r^2 \rangle^{A^{\prime}} - \langle r^2 \rangle^{A},
\label{IFS2}
\end{equation}
where $\delta\langle r^2 \rangle^{A,A^{\prime}}$ is the difference in nuclear
rms charge radii between isotopes $A$ and $A^{\prime}$, sometimes indicated by
the notation $A^{\prime} \leftarrow A$. A tabulation of nuclear mean-square
charge radii is available \cite{Angeli:1999}, but the results are not to very
high precision.

The quantity $\langle r^2 \rangle^{A}$ usually shows an odd-even staggering (a
``saw-tooth'' pattern), with exceptionally small values occurring for compact
nuclei with neutron magic numbers 20, 28, 50, 82, and 126 \cite{Aadndt1987}. It
is for this reason that magic-number isotopes such as $^{39}$K, $^{87}$Rb, and
$^{88}$Sr are often chosen as reference isotopes when tabulating values of
$\delta\langle r^2 \rangle^{A,A^{\prime}}$ \cite{AMadndt2013}. However, we do
not uniformly follow this convention here. In order to maintain consistency
with work performed by other authors, we deviate from it for Li and Rb, where
$^7$Li and $^{85}$Rb are chosen as reference isotopes.

Tiemann {\em et al.}\ \cite{Tiemann:1982} investigated non-Born-Oppenheimer
effects in the rotational spectra of group III/VII and IV/VI diatomic
molecules. They found good agreement with Watson's expressions, but with
anomalously large corrections for heavy atoms (Tl, Pb). Schlembach and Tiemann
\cite{Schlembach:1982} subsequently showed that the anomalous values can be
attributed to nuclear field shifts, and that these are the dominant
isotope-dependent effect for these species. For the vibronic spectra of PbS,
Kn\"ockel and Tiemann \cite{Knoeckel:1982} found that the field shift by itself
could explain the isotope dependence and terms due to mass shifts were
negligible. In a subsequent reevaluation of the experimental results,
Kn{\"o}ckel, Kr{\"o}ckertskothen and Tiemann \cite{KKTcp1985} revised the
magnitude of the field shift downwards substantially, but retained the overall
conclusion that field shifts dominate adiabatic corrections in the rotational
spectra of heavy-atom systems.

In the molecular case, the field shift for the free atoms can again be absorbed
into the zero of energy for each isotopolog, but its $R$-dependence contributes
to molecular binding energies and level spacings. For rotational spectra, the
key quantity is the derivative of the contact density with respect to $R$,
evaluated near the equilibrium geometry. Cooke {\em et al.}\ have determined
Dunham-type parameters from the rotational spectra of a wide variety of
diatomic molecules containing heavy elements, and identified large
Born-Oppenheimer breakdown effects that they interpreted as isotopic field
shifts \cite{CGpccp2004, CMGjms2004, CGCcp2004, CKGpccp2005, EDCcp2007,
SPDcpl2007, GBCpccp2008, DEGjms2008, KCRcjp2009, GBSjcp2011}. For a subset of
these studies they used density-functional theory with scalar relativistic
corrections to calculate contact density derivatives, and found reasonable
agreement with the experiments. However, Knecht and Saue \cite{KScp2012} have
carried out 4-component relativistic calculations on the TlI, PbTe, and PbS
systems; they obtained substantial disagreement with both the results of Cooke
{\em et al.}\ and the experiments, and questioned whether Cooke {\em et al.}\
had actually included relativistic corrections in Refs.\ \cite{CMGjms2004,
CGCcp2004, EDCcp2007}.


Expectation value and derivative approaches for obtaining approximate contact
densities are currently under development within the vibrational and
M{\"o}ss\-bauer spectroscopy communities
\cite{FRsb2012,Fccr2009,KFjctc2008,Fjcp2007,NHic2006,Nica2002,
LHLjacs2002,LLLjacs2001,MTams1985,NPvDprb1978,dVTRjcp1975,Srmp1964}. Within the
context of such calculations it has been shown that use of a relativistic
Hamiltonian is essential for obtaining accurate contact densities and their
geometry dependence for heavy nuclei
\cite{Fjcp2007,Fccr2009,KFvMtca2011,KScp2012,FZCjctc2012}. Incorporating
relativity directly into the electronic Hamiltonian can be done in various ways
(see, e.g., Refs.\ \cite{Lmp2010,Scpc2011,Pcr2012,Ajcp2012,Lpr2014} for recent
reviews and perspective articles).

When working within a fully relativistic framework, the use of finite-size
nuclear models is necessary to avoid singularities in the wavefunction.
Finite-size nuclear models may seem at first problematic since the perturbation
theory approach for obtaining contact densities outlined above assumes the
point-nucleus approximation as a zeroth-order starting point. A more accurate
method for obtaining field shifts involves integration of the electron density
over the nuclear volume. It has been shown in Refs.\ \cite{KFvMtca2011} and
\cite{KScp2012} that the error introduced by replacing such integration by the
finite-nucleus contact density is on the order of 10\% for absolute contact
densities, with smaller errors for changes with respect to geometry.

Many additional aspects of the computational methods for contact densities have
been considered in Refs.\ \cite{MLRcpl2008,MWRjcp2010,KFvMtca2011}. The
importance of electron correlation has been examined in Ref.\
\cite{KFvMtca2011}, where it was shown for mercury fluoride systems that
correlation effects at the CCSD level contribute as much as $\sim20$\% to
bonding-induced changes in contact densities. However, the resulting
bonding-induced changes were no more accurate than those calculated at the HF
level. Including a perturbative correction for triple excitations was found to
affect the bonding-induced changes by $\sim5$\%. Inclusion of core-valence
correlation was also shown to be important, contributing at the same level as
the perturbative triples correction. When instead density-functional theory
(DFT) methods were considered, multiple studies concluded that DFT contact
densities are of comparable accuracy to HF
\cite{MWRjcp2010,KFjctc2008,KFvMtca2011}, though hybrid functionals with HF
exchange were shown to give better results than pure functionals
\cite{KFjctc2008}. It was also shown in Ref.\ \cite{MWRjcp2010} that much more
accurate results were produced by using basis sets in their fully uncontracted
forms.

\section{Results and discussion}
\label{sec:RandD}
\subsection{Benchmarking computed isotopic mass shifts}
\label{sec:b_ims}

There are not many molecular electronic structure packages that currently have
the capability to calculate mass shifts (i.e.\ adiabatic or diagonal
Born-Oppenheimer corrections) at the coupled-cluster level. In the present work
we compute isotopic mass shifts for atomic and molecular species using the DBOC
facility in the CFour package \cite{CFour}. Core orbitals are correlated in all
calculations, since core-valence correlation has been shown to contribute
significantly to DBOCs \cite{GTKjcp2006}. Parallelized analytic derivative
codes are used.

For all atoms except H, DBOC calculations were initially performed using small
basis sets, in particular DZP for the alkali metals and Sr
\cite{DZP1,DZP2,DZP3,DZP4} and the WTBS basis set for Yb \cite{WTBS1,WTBS2}.
The larger ANO-RCC basis sets \cite{ANORCC1,ANORCC2} were also employed in some
cases. The ANO-RCC basis sets have previously been shown to provide excellent
dissociation energies, geometrical parameters, and electric properties for
systems involving heavy elements, as demonstrated, for example, in a recent
study on LiCs \cite{SFOjpb2009}. While these basis sets are generally
recommended only for relativistic calculations, the derivative programs
required for the evaluation of DBOCs within a relativistic framework are not
available in the current public release of CFour. Employing basis sets designed
for relativistic calculations in non-relativistic work is somewhat
questionable. However, relativistic effects are significant only for heavier
elements, where DBOCs are expected to become small compared to the field shift.
Our overall conclusions regarding DBOCs are thus unlikely to be affected by the
errors due to basis-set incompleteness and neglect of relativity.

CFour allows the use of both spin-restricted HF (RHF) and spin-unrestricted HF
(UHF) references. Unrestricted methods seem at first sight an appealing choice,
as they offer a better description of the highly stretched molecule near
dissociation and are directly applicable to individual doublet atomic species
at dissociation. However, they may find wavefunctions that have a lower
symmetry than the nuclear framework, and such symmetry-broken solutions are
also often spin-contaminated to some degree
\cite{Stanton1994,Krylov2000,Schlegel2002,Schlegel2009}. This may cause
additional complications in the evaluation of second-order properties such as
DBOCs. We explore such phenomena in detail in the following section. Where UHF
methods were employed, the lowest-energy UHF eigenstate was located within a
reduced computational symmetry (C$_{\mathrm s}$) by following the appropriate
eigenvalue of the orbital rotation Hessian matrix from the totally symmetric
RHF solution to the symmetry-broken solution.

Before computing the molecular (bonding-induced) isotopic mass shifts of
interest in this work, we first consider whether molecular calculations with
CFour can yield accurate absolute values of {\em atomic} mass shifts. Here
spin-restricted calculations were performed on the corresponding homonuclear
diatomic system at large values of the internuclear distance. For the
$^6$Li$\leftarrow$$^7$Li mass shift, DBOCs were computed at the
RHF-CCSD/ANO-RCC and UHF-CCSD/ANO-RCC levels, resulting in values (per atom) of
$-22.23$ cm$^{-1}$ and $-22.25$ cm$^{-1}$, respectively. These compare
favorably with the atomic physics literature value of $-21.36$ cm$^{-1}$,
obtained by applying the appropriate Rydberg factors (see, e.g., King's
description in Ref. \cite{Kjms1997}) to the near-exact total energy of Li
calculated by Puchalski and Pachucki \cite{PPpra2006}.

\begin{figure*}
\centering
\begin{minipage}{\textwidth}
\subfigure[]{
        \resizebox*{.48\textwidth}{!}{
                \fbox{\includegraphics[width=.99\textwidth]{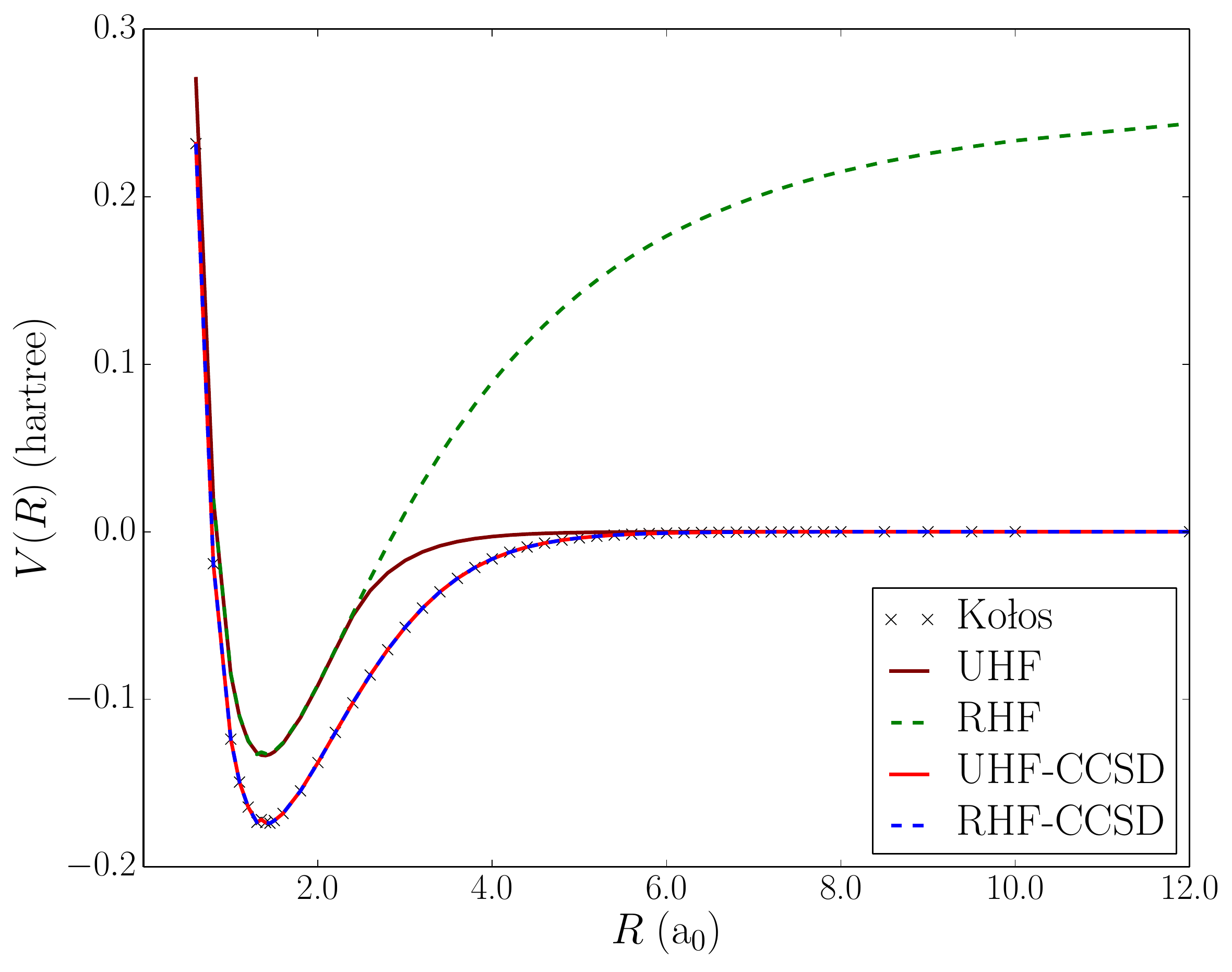}}}
        \label{fig:H2meth:a}}
\subfigure[]{
        \resizebox*{.48\textwidth}{!}{
                \fbox{\includegraphics[width=.99\textwidth]{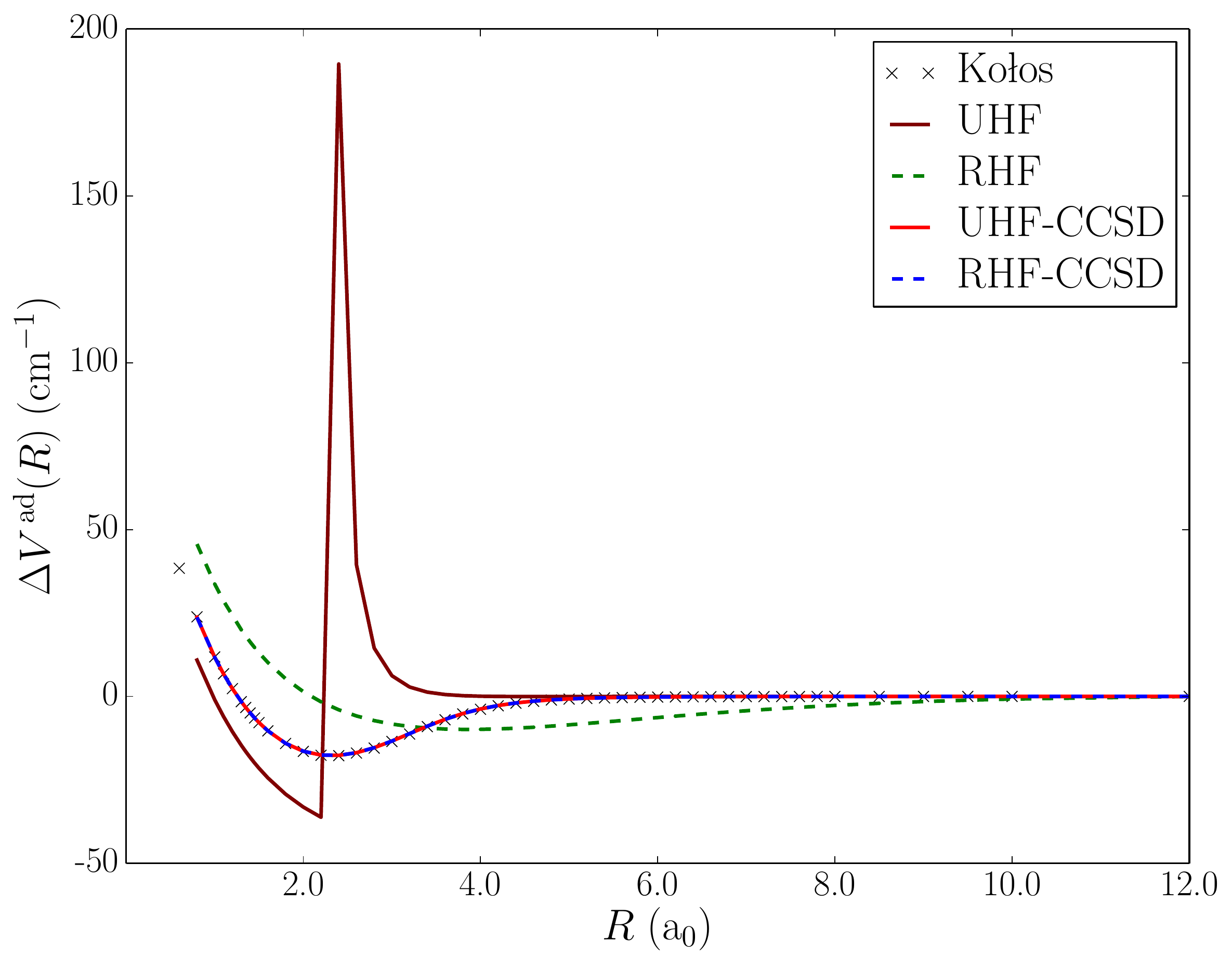}}}
        \label{fig:H2meth:b}}
\caption{(a) Potential curves for H$_2$, computed using the ANO-RCC basis set
with various electronic structure methods. (b) The corresponding DBOC functions
$\Delta V^{\rm ad}(R)$ for $^1$H$^1$H. \label{fig:H2meth}}
\end{minipage}
\end{figure*}

Next we investigate the reliability of restricted and unrestricted references
for the computation of molecular DBOCs. We first consider H$_2$, for which the
nearly exact energies and adiabatic corrections by Ko{\l}os and coworkers are
available for comparison \cite{Kolos:1964, Kolos:1965, KRjcp1993}. Various
approximate potential energy curves and absolute DBOC functions for $^1$H$^1$H
are shown in Fig.\ \ref{fig:H2meth}. All potential energies are shown with
respect to the exact asymptotic value, 0.5 Hartree, and the DBOC values are
shown relative to an asymptotic value obtained from a molecular calculation at
large $R$. The behavior of the H$_2$ potential curves as described by the RHF,
UHF, RHF-CCSD, and UHF-CCSD methods is discussed in elementary textbooks and we
include them in Fig.\ \ref{fig:H2meth:a} only to contrast their characteristics
with the corresponding DBOC functions.

The computed DBOC functions $\Delta V^{\rm ad}(R)$ for $^1$H$^1$H are shown in
Fig.\ \ref{fig:H2meth:b}. The UHF function is qualitatively wrong, exhibiting
an unphysical pole-like feature near 2.2 bohr, which we discuss in detail
below. The RHF function is smooth, but in poor quantitative agreement with the
results of Ko{\l}os and Rychlewski \cite{KRjcp1993}. However, the RHF-CCSD and
UHF-CCSD methods are exact for this system, except for basis-set incompleteness,
and exhibit errors in DBOCs no larger than 1\%.

\begin{figure*}
\begin{center}
\begin{minipage}{\textwidth}
\subfigure[]{
        \resizebox*{.48\textwidth}{!}{\includegraphics{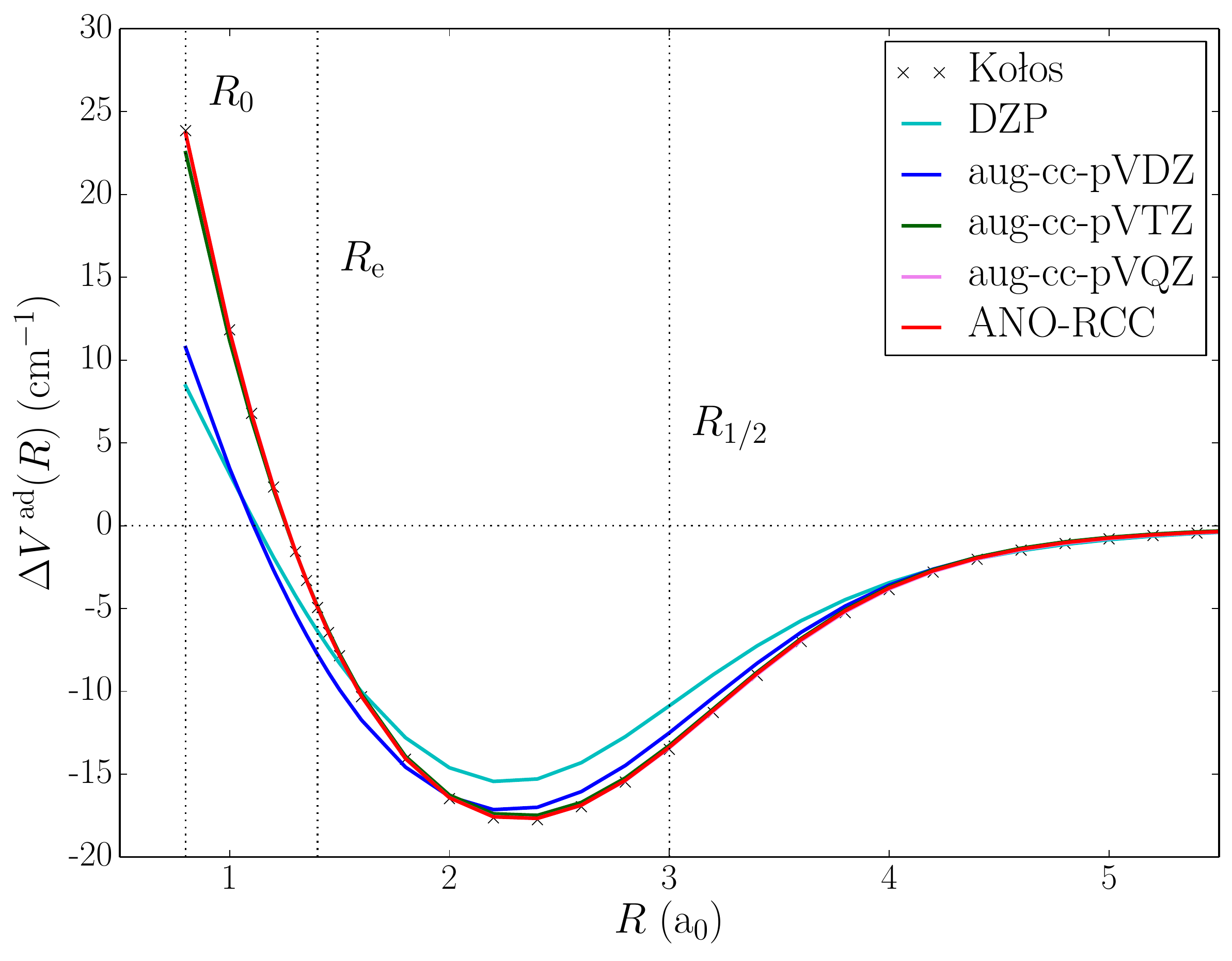}}
        \label{fig:basDBOC:a}}
\subfigure[]{
        \resizebox*{.48\textwidth}{!}{\includegraphics{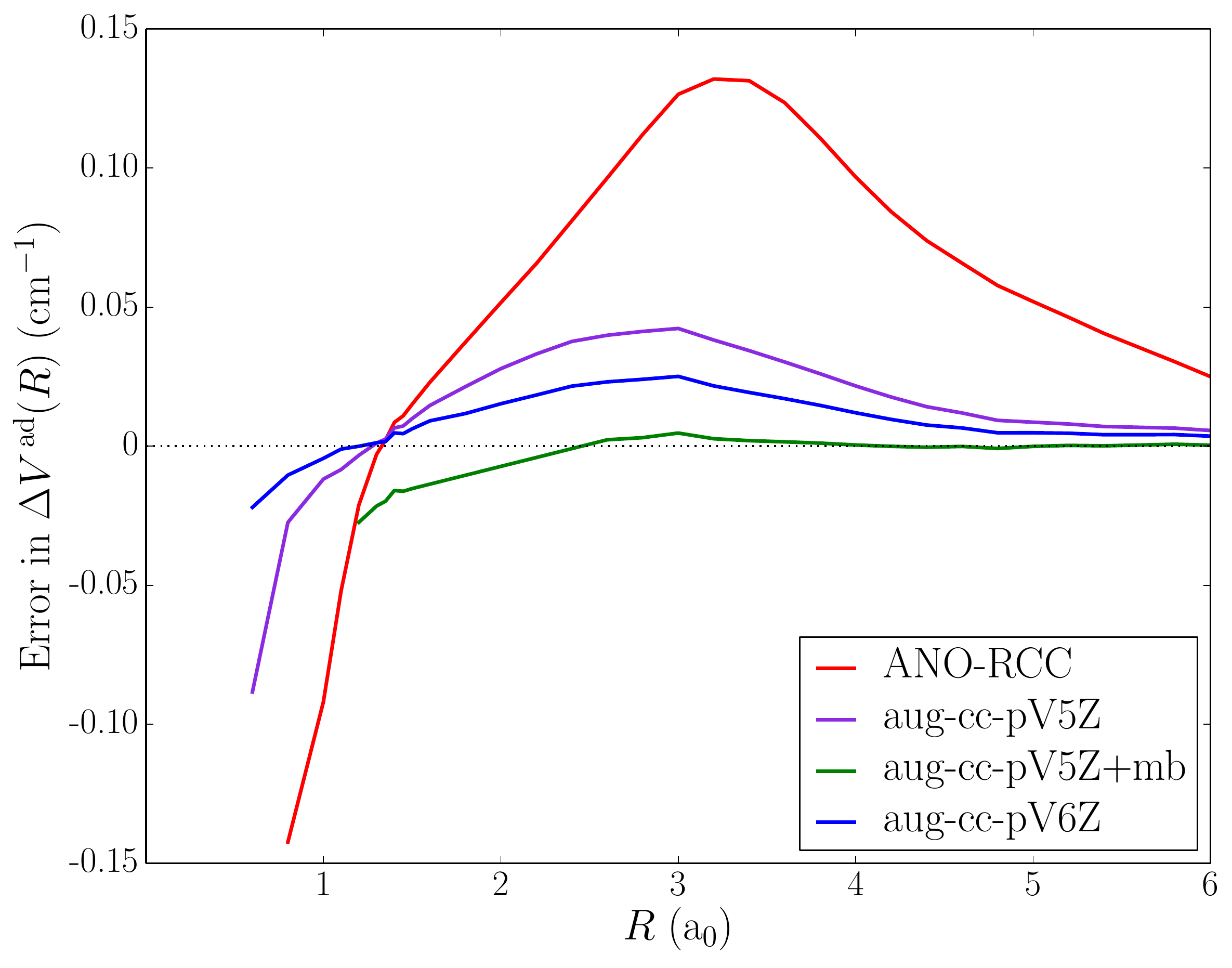}}
        \label{fig:basDBOC:b}}
\end{minipage}
\caption{(a) DBOC functions $\Delta V^{\rm ad}(R)$ for $^1$H$^1$H, computed at
the CCSD level with various basis sets. (b) Errors in DBOC functions computed
with CCSD and various basis sets taken with respect to the near-exact values of
Ko{\l}os and Rychlewski \cite{KRjcp1993}. \label{fig:basDBOC}}
\end{center}
\end{figure*}

Figure \ref{fig:basDBOC:a} shows the sensitivity of the computed CCSD DBOC
function to the quality of the basis set. Vertical lines mark the radial
position of key points on the potential curve. These include the inner turning
point at dissociation $R_0$ (so that $V(R_0) = 0$), the distance
$R_{\mathrm{e}}$ at the potential minimum, and the point $R_{1/2}$ where the
energy is half way between the minimum and dissociation
($V(R_{1/2})=\frac{1}{2} V(R_{\mathrm{e}})$). On the scale of the plot, the
ANO-RCC and aug-cc-pVQZ basis set curves lie directly on top of the results of
Ko{\l}os and Rychlewski \cite{KRjcp1993}. Even the much smaller and more
affordable DZP basis set gives results that are accurate within $\sim$10\%.
Because of this reasonable accuracy and its availability for most elements
across the periodic table, the DZP basis set will be heavily utilized in this
work for exploring trends in isotope shifts with atomic number.

To demonstrate the level of accuracy possible for DBOCs obtained with large
basis sets, Fig.\ \ref{fig:basDBOC:b} shows the errors in the computed DBOC
functions with respect to the results of Ko{\l}os and Rychlewski
\cite{KRjcp1993}. The ANO-RCC basis set gives errors that do not exceed 0.13
cm$^{-1}$ at distances larger than the inner turning point ($R>0.8$ bohr). The
ANO-RCC basis set is known to perform similarly to the aug-cc-pVQZ basis set,
so we also computed DBOC functions with the aug-cc-pV5Z and aug-cc-pV6Z basis
sets, where maximum errors were found to drop to 0.042 and 0.025 cm$^{-1}$,
respectively. The addition of midbond functions (designated in the figure
legend as ``+mb'') was also investigated and they were found to reduce errors
greatly for $R>2.0$ bohr, while increasing errors somewhat for $R<2.0$ bohr.
The performance of all basis sets tested here degrades rapidly in the region
$R<1.0$ bohr, probably because modern basis sets are tuned for optimum
performance near the equilibrium bond length ($R_{\rm e}=1.4$~bohr).

The unphysical pole-like feature in the UHF function in Fig.\
\ref{fig:H2meth:b} is analogous to singularities that have been studied in the
context of other properties including quadratic force constants
\cite{CSASjcp1997} and indirect nuclear spin-spin coupling constants
\cite{AGcp2009}. Such poles arise because the second derivatives of the
correlated energies depend upon the orbital rotation parameters, which
themselves are not continuously differentiable through the region of the
transition from a symmetry-conserved to a symmetry-broken wavefunction. This
phenomenon is sometimes referred to as an orbital instability envelope
\cite{CSASjcp1997}.

\begin{figure*}
\centering
\begin{minipage}{\textwidth}
\subfigure[]{
        \resizebox*{.48\textwidth}{!}{
                \fbox{\includegraphics[width=.99\textwidth]{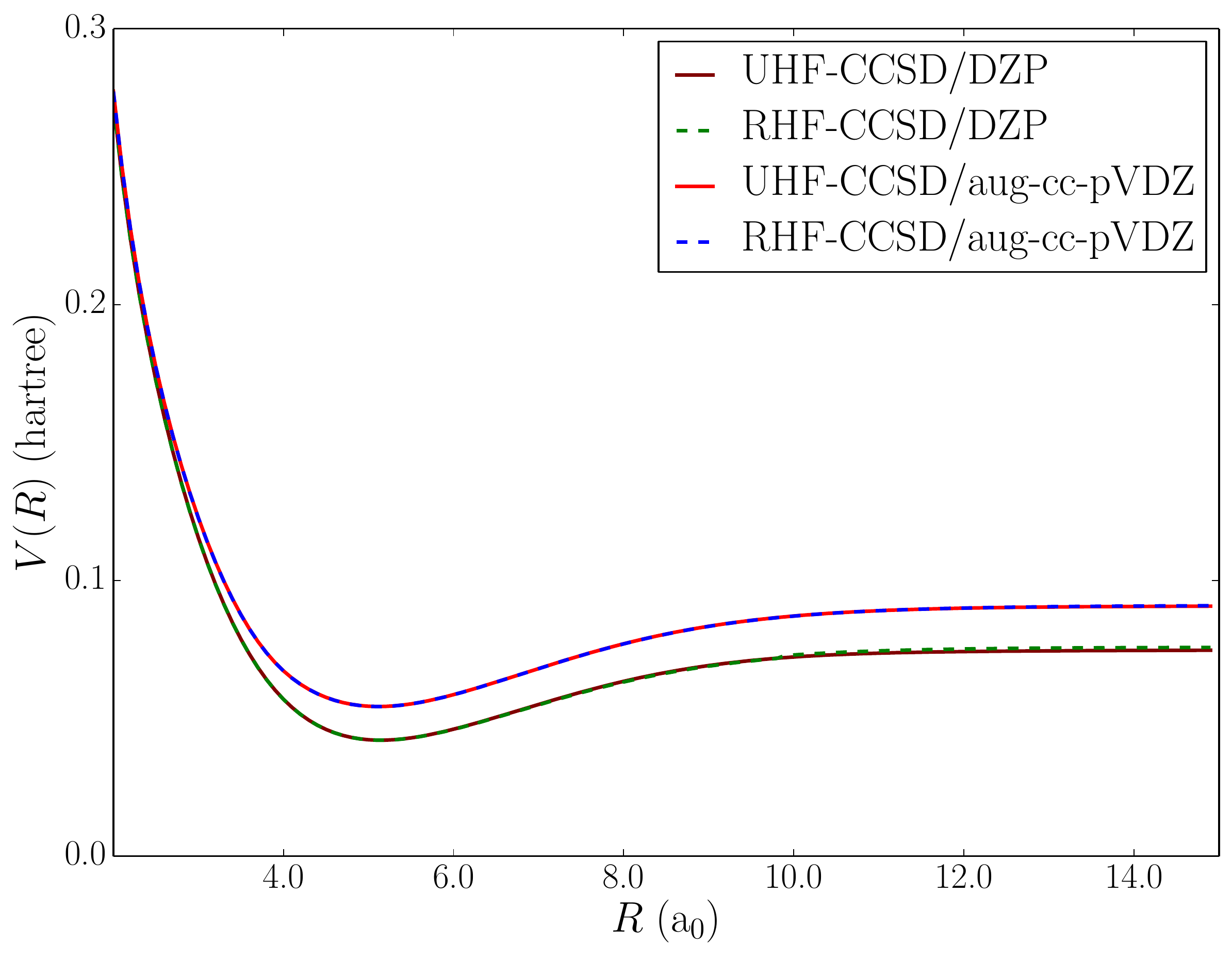}}}
        \label{fig:li2meth:a}}
\subfigure[]{
        \resizebox*{.48\textwidth}{!}{
                \fbox{\includegraphics[width=.99\textwidth]{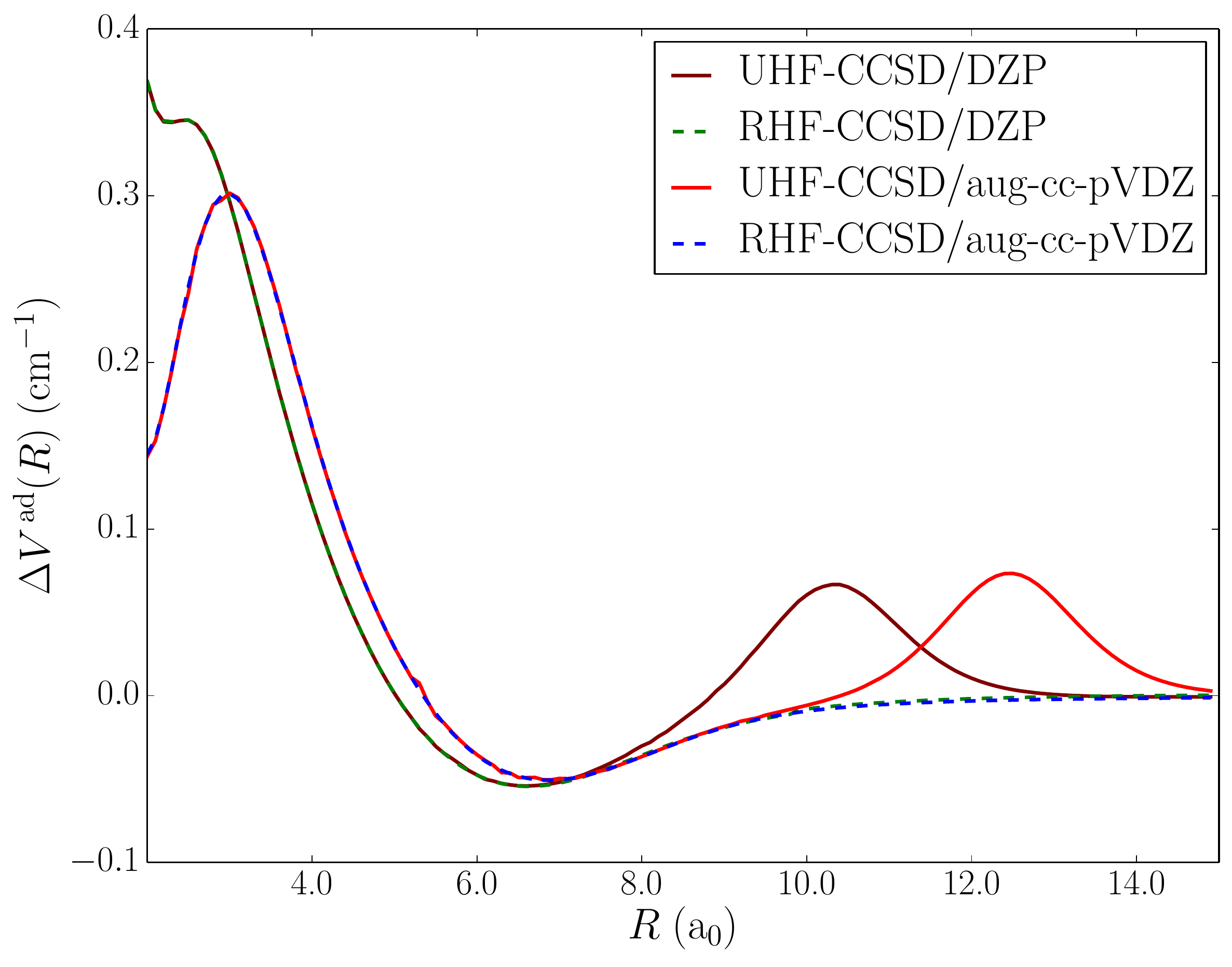}}}
        \label{fig:li2meth:b}}
\caption{(a) Potential curves for Li$_2$ from RHF-CCSD and UHF-CCSD
calculations using the DZP and aug-cc-pVDZ basis sets. (b) The corresponding DBOC
functions $\Delta V^{\rm ad}(R)$ for $^7$Li$^7$Li. \label{fig:li2meth}}
\end{minipage}
\end{figure*}

UHF-CCSD may also produce orbital instability envelope artifacts in molecules
with more than 2 electrons. This is demonstrated in Fig.\ \ref{fig:li2meth},
which shows Li$_2$ potential energy curves and DBOCs computed using RHF-CCSD
and UHF-CCSD with the DZP and aug-cc-pVDZ basis sets. While the RHF-CCSD and UHF-CCSD
potentials are virtually identical for each basis set, the corresponding DBOC
functions are not. For both basis sets, the UHF-CCSD results exhibit an
unphysical peak in a region where the DBOC radial function should
asymptotically approach zero. For this reason we choose to use RHF-CCSD
calculations in preference to UHF-CCSD calculations of DBOCs in the following
sections.

\subsection{Isotopic mass shifts for diatomic molecules topical in ultracold physics}
\label{sec:r_ims}

\begin{figure*}
\begin{center}
\begin{minipage}{\textwidth}
\subfigure[]{
        \resizebox*{.48\textwidth}{!}{\includegraphics{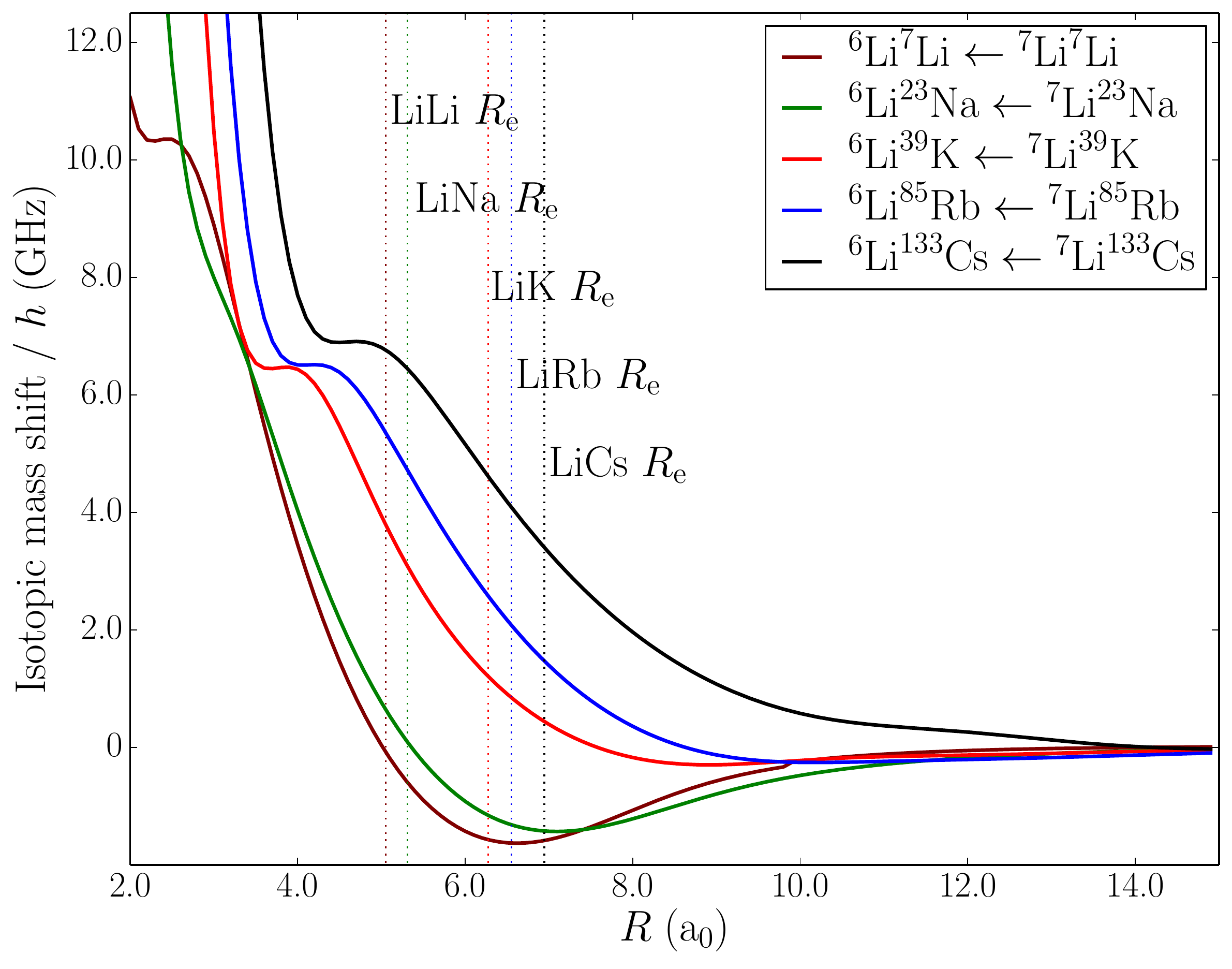}}
        \label{fig:LiXDBOC}}
\subfigure[]{
        \resizebox*{.48\textwidth}{!}{\includegraphics{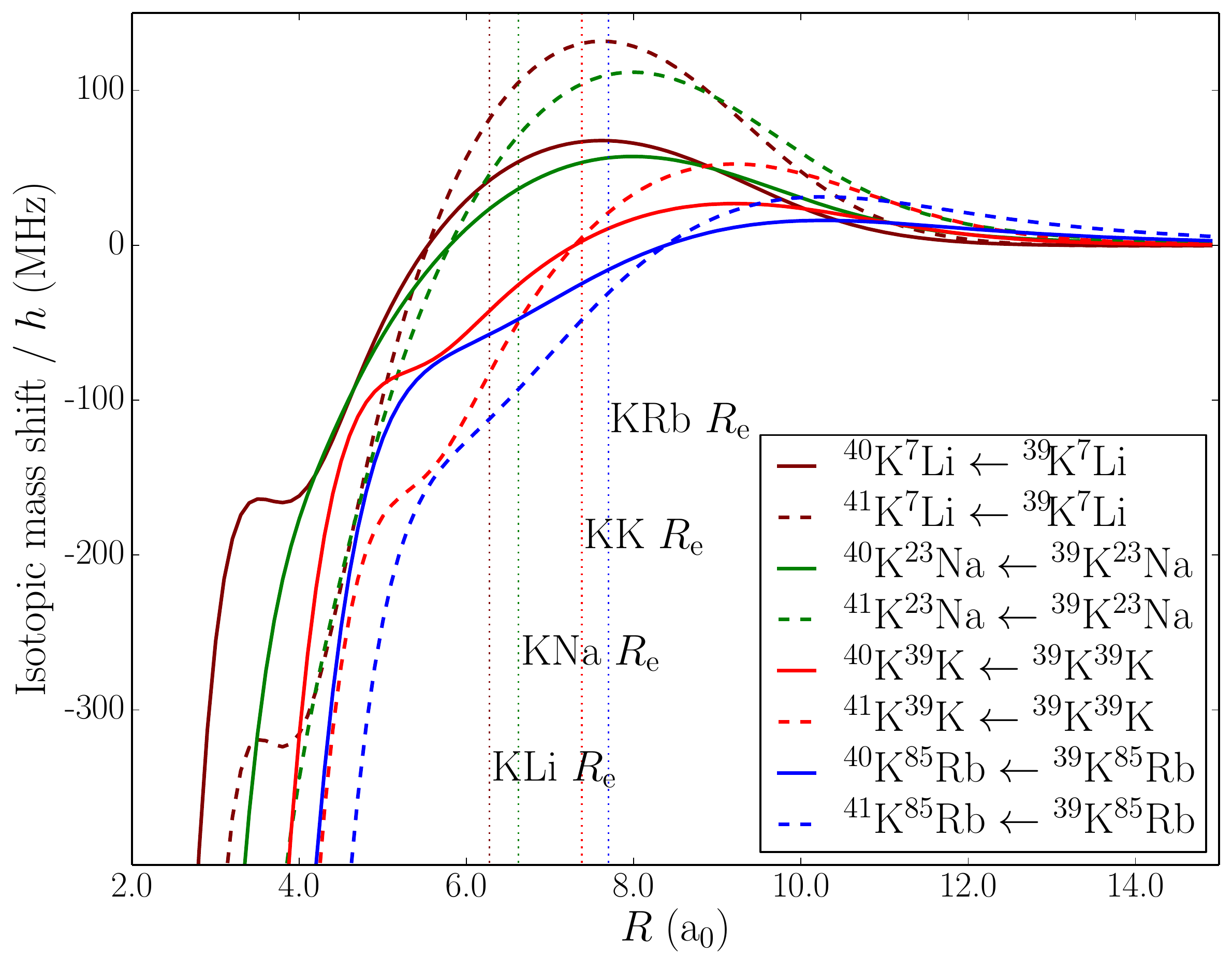}}
        \label{fig:KXDBOC}}
\\
\subfigure[]{
        \resizebox*{.48\textwidth}{!}{\includegraphics{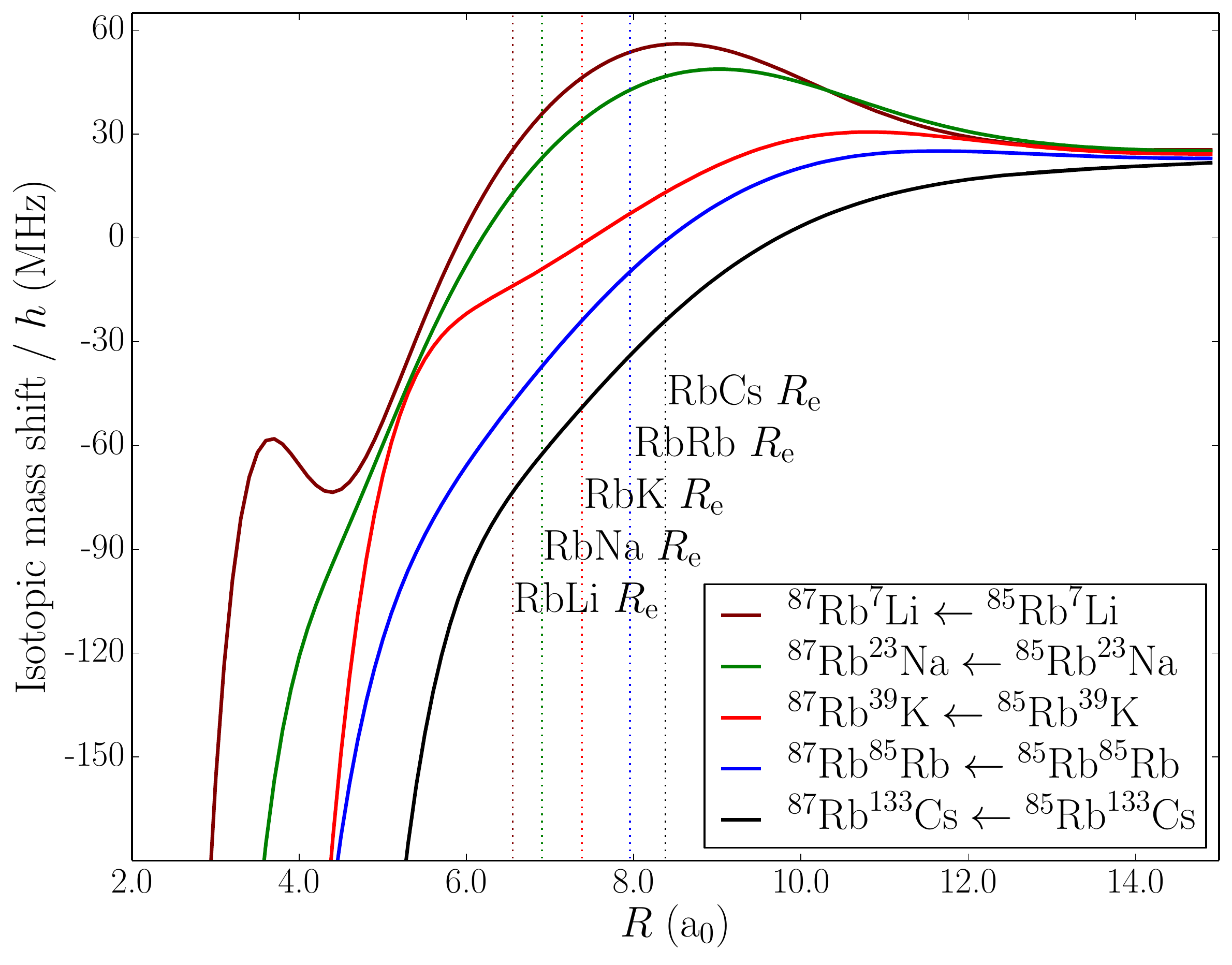}}
        \label{fig:RbXDBOC}}
\subfigure[]{
        \resizebox*{.48\textwidth}{!}{\includegraphics{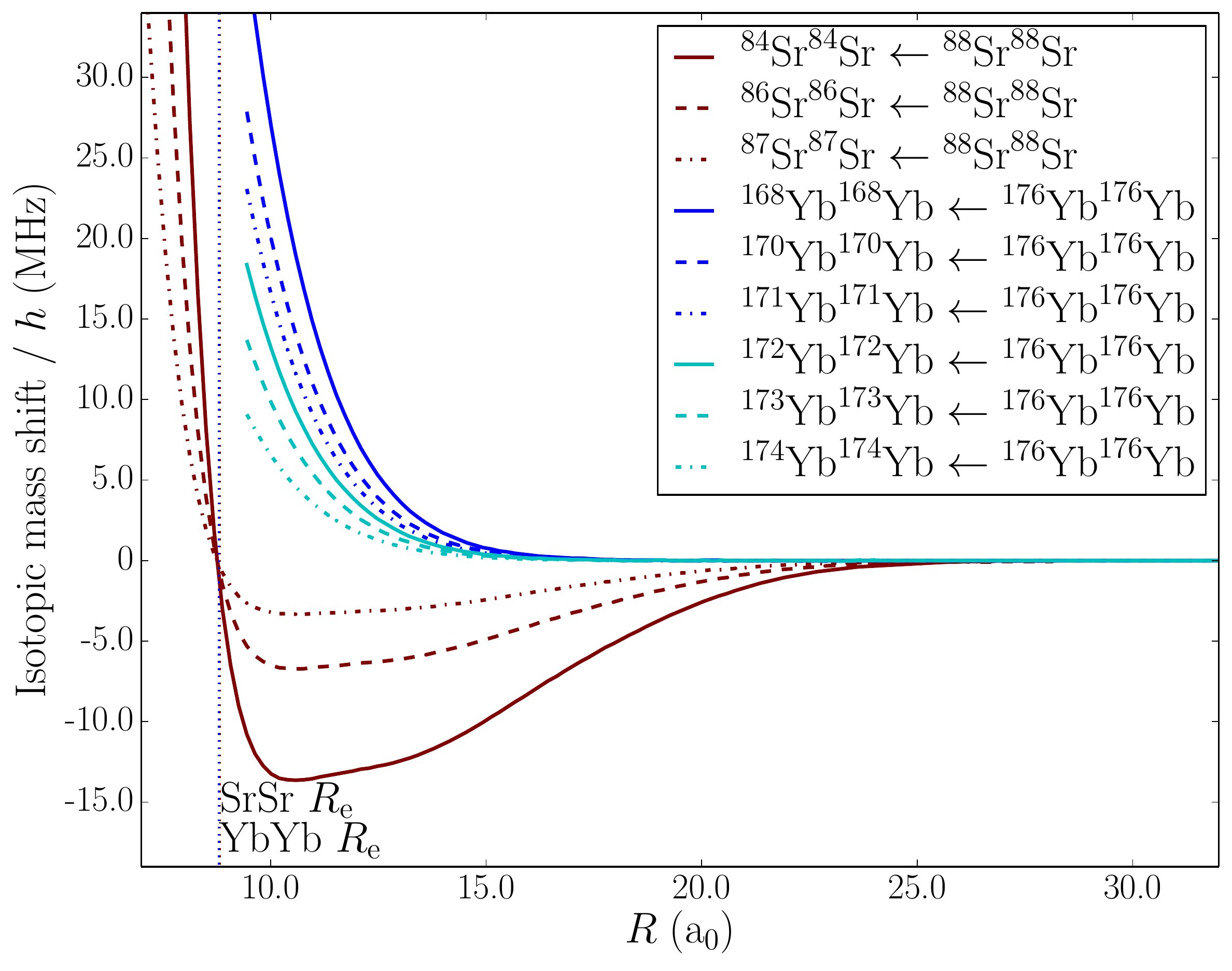}}
        \label{fig:SrYbDBOC}}
\end{minipage}
\caption{Bonding-induced changes in mass shifts $\delta V^{\rm ad}(R)$ (a) Li in
alkali-metal dimers; (b) K in alkali-metal dimers; (c) Rb in alkali-metal dimers;
(d) homonuclear isotopologs of Sr$_2$ and Yb$_2$.
\label{fig:DBOCs}}
\end{center}
\end{figure*}

Radial isotopic mass-shift functions, obtained from DBOCs computed at the
RHF-CCSD/DZP level of theory, are shown in Fig.\ \ref{fig:DBOCs} for the ground
states of all the alkali-metal dimers and the molecules Sr$_2$ and Yb$_2$.
These are all molecules of interest in the field of ultracold molecules. The
mass shifts are defined as differences
\begin{equation}
\delta V_{\beta\leftarrow\alpha}^{\rm ad}(R)=\Delta V_\beta^{\rm ad}-\Delta V_\alpha^{\rm ad}
\end{equation}
between the adiabatic corrections for isotopologs $\beta$ and $\alpha$. The
radial positions $R_{\rm e}$ of the corresponding potential minima are
indicated by labeled and color-coded vertical lines \cite{FDVjcp2014, Kjcp2008,
MPTZRijqc2009}. The qualitative shape of the curves in Figs.\ \ref{fig:LiXDBOC}
and \ref{fig:SrYbDBOC} is inverted compared to those in Figs.\ \ref{fig:KXDBOC}
and \ref{fig:RbXDBOC}, but this is simply because for Li, Sr and Yb the
reference isotope is the heaviest, while for K and Rb it is the lightest.

We first consider $^6$Li$\leftarrow$$^7$Li mass shifts in alkali-metal
diatomics containing Li, which are shown in Figure \ref{fig:LiXDBOC}. The
functions for Li$_2$ and LiNa have the same general shape as for H$_2$, with a
negative segment at long range and a positive segment at short range. The
zero-crossing is close to $R_{\rm e}$ for Li$_2$ and LiNa, but moves outwards
faster than $R_{\rm e}$ for the heavier alkali metals, and for LiCs the
function does not cross zero until about 14 bohr. This may well be due to the
increasing contribution of charge transfer to the bonding in LiK, LiRb and
LiCs, even at 10 to 15 bohr, as evidenced by their dipole-moment functions
\cite{Aymar:2005}. In Section \ref{sec:emp} we will investigate further whether
these mass-shift functions are in agreement with spectroscopically derived
isotope shifts. It is worth noting that some mass-shift functions cross zero so
close to their respective potential minima that reporting mass shifts at
$R_{\rm e}$ or at any other single point on the curve can be uninformative.

Isotopic mass shifts from DBOC calculations at the RHF-CCSD/DZP level are shown
for isotopologs of alkali-metal diatomics involving K (with $^{39}$K as
reference) in Fig.\ \ref{fig:KXDBOC} and for those involving Rb (with $^{85}$Rb
as reference) in Fig.\ \ref{fig:RbXDBOC}. As expected, the mass shifts for
$^{41}$K are about twice those for $^{40}$K, with the exact ratio determined by
the changes in the isotopic mass. KCs curves are omitted from Fig.\
\ref{fig:KXDBOC} because many of the corresponding RHF calculations did not
converge. The overall magnitude of the mass shift for
$^{85}$Rb$^{85}$Rb$\leftarrow$$^{85}$Rb$^{87}$Rb substitution, $\delta
V^{\mathrm{ad}}(R_{\mathrm e})\approx-30$ MHz, is consistent with the
order-of-magnitude absolute-value estimate suggested in Ref.\ \cite{VKKpra2009}
for the correction due to Born-Oppenheimer breakdown.

Figure \ref{fig:SrYbDBOC} shows mass shifts from DBOC calculations at the
RHF-CCSD/DZP and RHF-CCSD/WTBS levels for homonuclear Sr$_2$ and Yb$_2$
diatomics, respectively, with $^{88}$Sr and $^{176}$Yb as reference isotopes.
Both Sr$_2$ and Yb$_2$ have mass shifts that are positive at short range (but
well outside the inner turning points, which are both between 7 and 7.5 bohr
\cite{Stein:2008, MPTZRijqc2009}). The mass shifts for Sr$_2$ have a
significant negative component at long range, which dominates near $R_{\rm e}$.
Yb$_2$, by contrast, has very weak mass shifts at long range, and the absolute
values are at least an order of magnitude smaller at the equilibrium bond
length than those for Sr$_2$ or any of the other systems shown in Figs.\
\ref{fig:KXDBOC} to \ref{fig:SrYbDBOC}. However, it should be noted that the
minimal WTBS basis set used for Yb is considerably less flexible than the DZP
basis set used for Sr, and is expected to produce too soft a short-range
repulsive interaction and provide a poorer description of long-range forces
\cite{Szabo}. These features are reflected in Fig.\ \ref{fig:SrYbDBOC}.
Unfortunately, DBOC calculations with the larger basis sets that are currently
available for Yb would be computationally very expensive. For now, computations
at this level of theory must suffice; they demonstrate that DBOCs of Sr$_2$ and
Yb$_2$ have similar qualitative features, with those for Yb$_2$ being much
smaller in magnitude.

\subsection{Isotopic field shifts for diatomic molecules topical in ultracold physics}
\label{sec:r_ifs}

Field shifts can also contribute significantly to total isotope shifts. The
field shift is approximately proportional to the contact density (Eq.\
\ref{IFS1}), so this is the key quantity to compute using electronic structure
theory. We have carried out extensive benchmark calculations on LiRb to compare
the results of different approaches using the DIRAC, MOLCAS and ADF packages,
which are described in the Supplementary Material. The packages all use
different treatments of relativity, and offer different options for including
electron correlation, either at the coupled-cluster level (DIRAC only) or using
density-functional theory (all three programs). Unfortunately the different
treatments offer less consistent results than we would have wished, and it is
clear that further work on the methods is needed to develop quantitatively
accurate procedures.

For the alkali-metal dimers and Sr$_2$ we have chosen to proceed with a
consistent set of full 4-component calculations of contact densities, using DFT
calculations with the B3LYP functional \cite{Beck93,Lee:1988}, which has a good
track record for calculating changes in contact densities due to chemical
bonding \cite{KFjctc2008}. These were performed with the DIRAC package
\cite{DIRAC11}, since it is the only program used in this work which can offer
a potentially exact treatment of relativity. The DIRAC calculations employed
the aug-cc-pVTZ-DK basis set for Li \cite{Dunning} and the v3z basis sets of
Dyall \cite{Djpca2009} for other elements. The basic Gaussian finite nuclear
charge distribution model, which has been shown to be sufficiently reliable to
yield bond-induced changes in contact densities \cite{MLRcpl2008}, was used
with parameters given by Visscher and Dyall \cite{VDadndt1997}. Contact
densities were obtained by evaluating the expectation value $\langle 0 |
\delta(r-R)|0\rangle$. An ultrafine grid was employed to ensure converged
results in the exchange-correlation evaluation.

For the Yb$_2$ system DIRAC failed to converge and ADF \cite{ADF1,ADF2} was
used instead to evaluate $\rho(0)$. In this case scalar relativistic effects
(which are the equivalent of Darwin and mass-velocity terms in the Breit-Pauli
Hamiltonian) were included via the zero-order regular approximation (ZORA)
\cite{CPDps1986, vLBSjcp1993, vLBSjcp1994, FSvLcpl1995}. The ZORA/QZ4P basis
set \cite{vLBjcc2003}, which is an all-electron basis sets of Slater-type
orbitals (STO), was used in conjunction with a point-charge nuclear model. ADF
parameters were chosen to enable use of the true (exact) electron density in
the exchange-correlation potential. As shown in the Supplementary Information,
bonding-induced changes in field shifts obtained by this method cannot be
regarded as quantitative, and might indeed be only order-of-magnitude
estimates, but even this is valuable in understanding which effects are
dominant for Yb$_2$.

\begin{figure}
\centering
\includegraphics[width=0.60\textwidth]{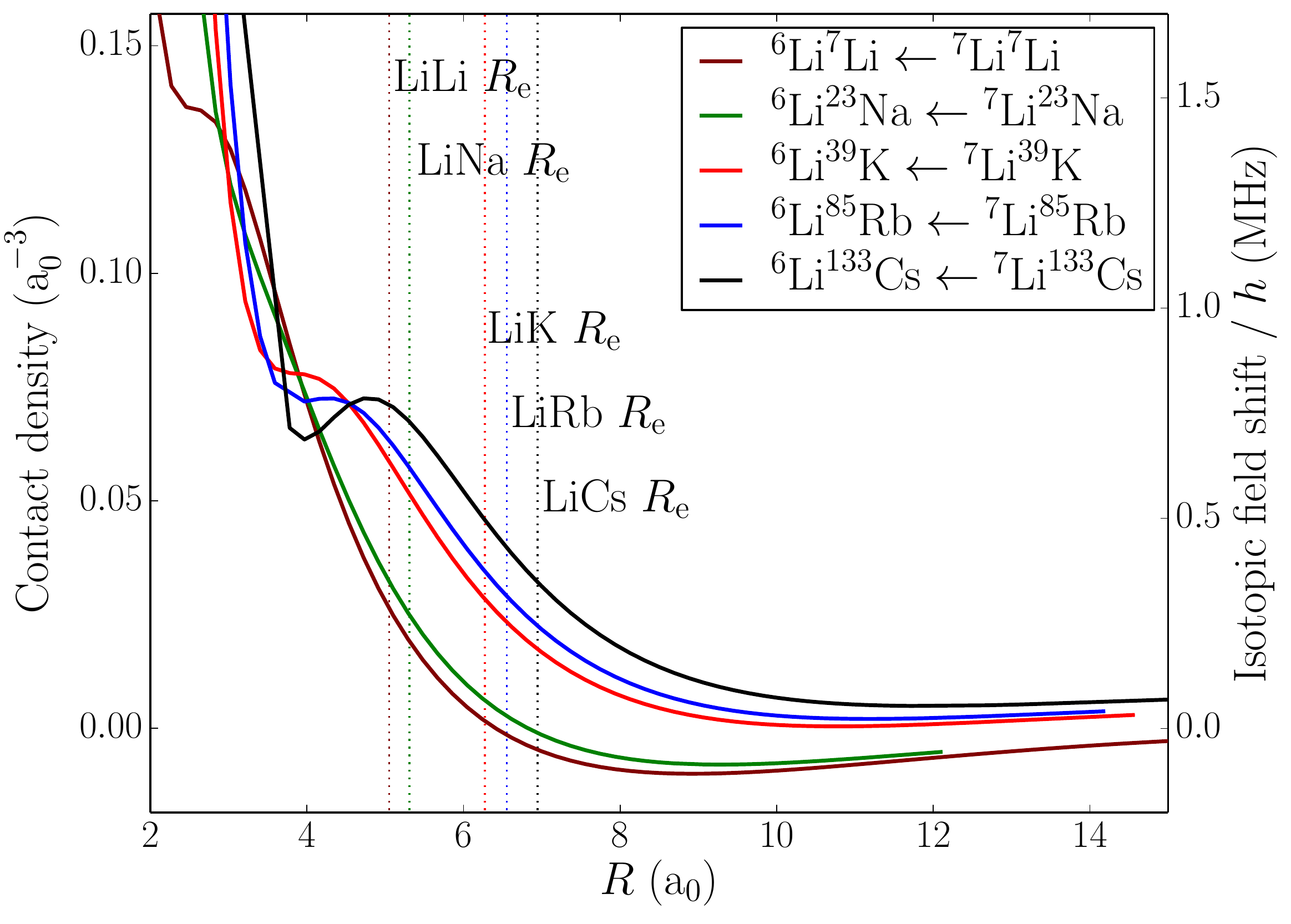} \label{fig:LiXFS} \\ (a) \\
\includegraphics[width=0.60\textwidth]{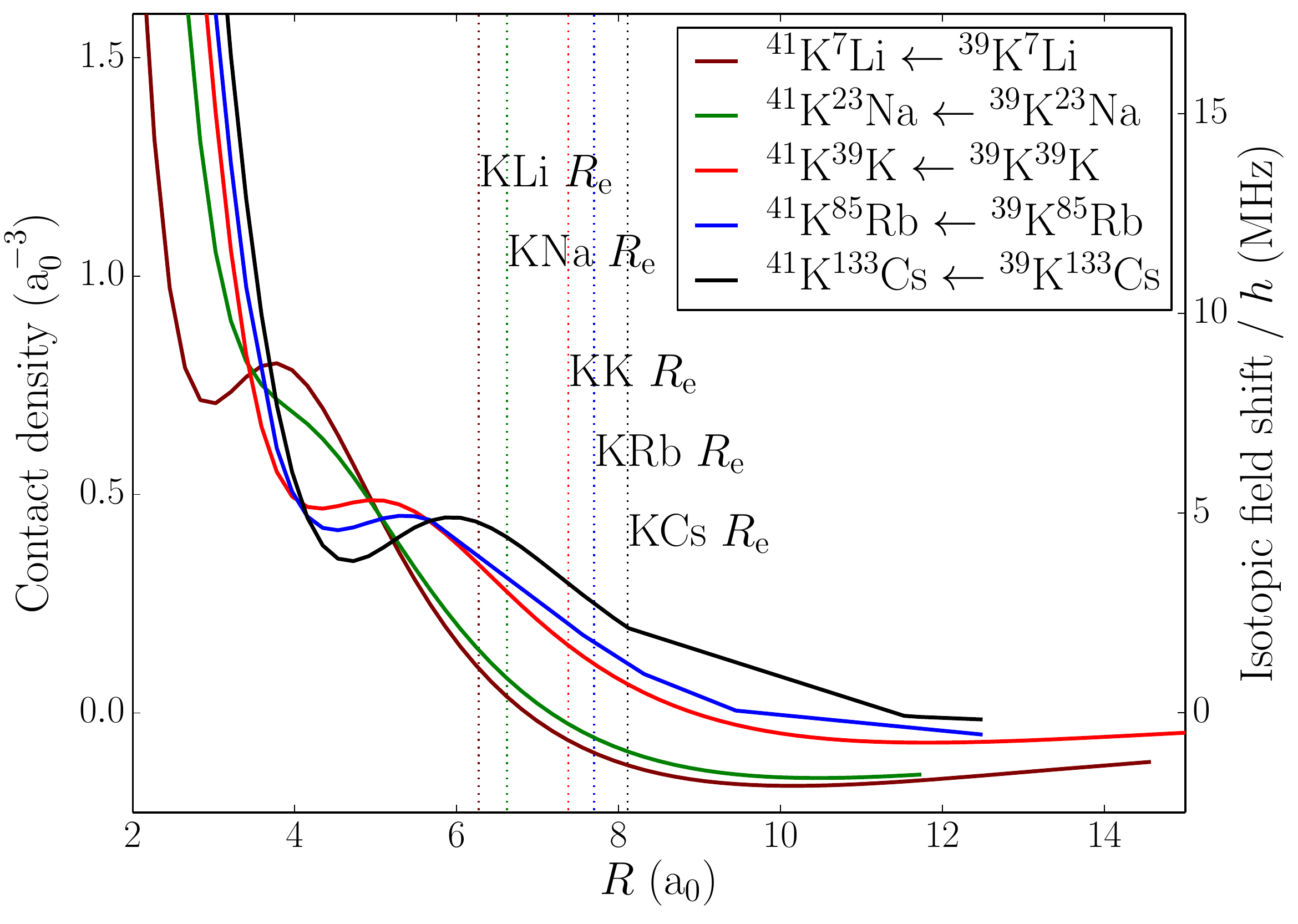} \label{fig:KXFS} \\ (b) \\
\includegraphics[width=0.60\textwidth]{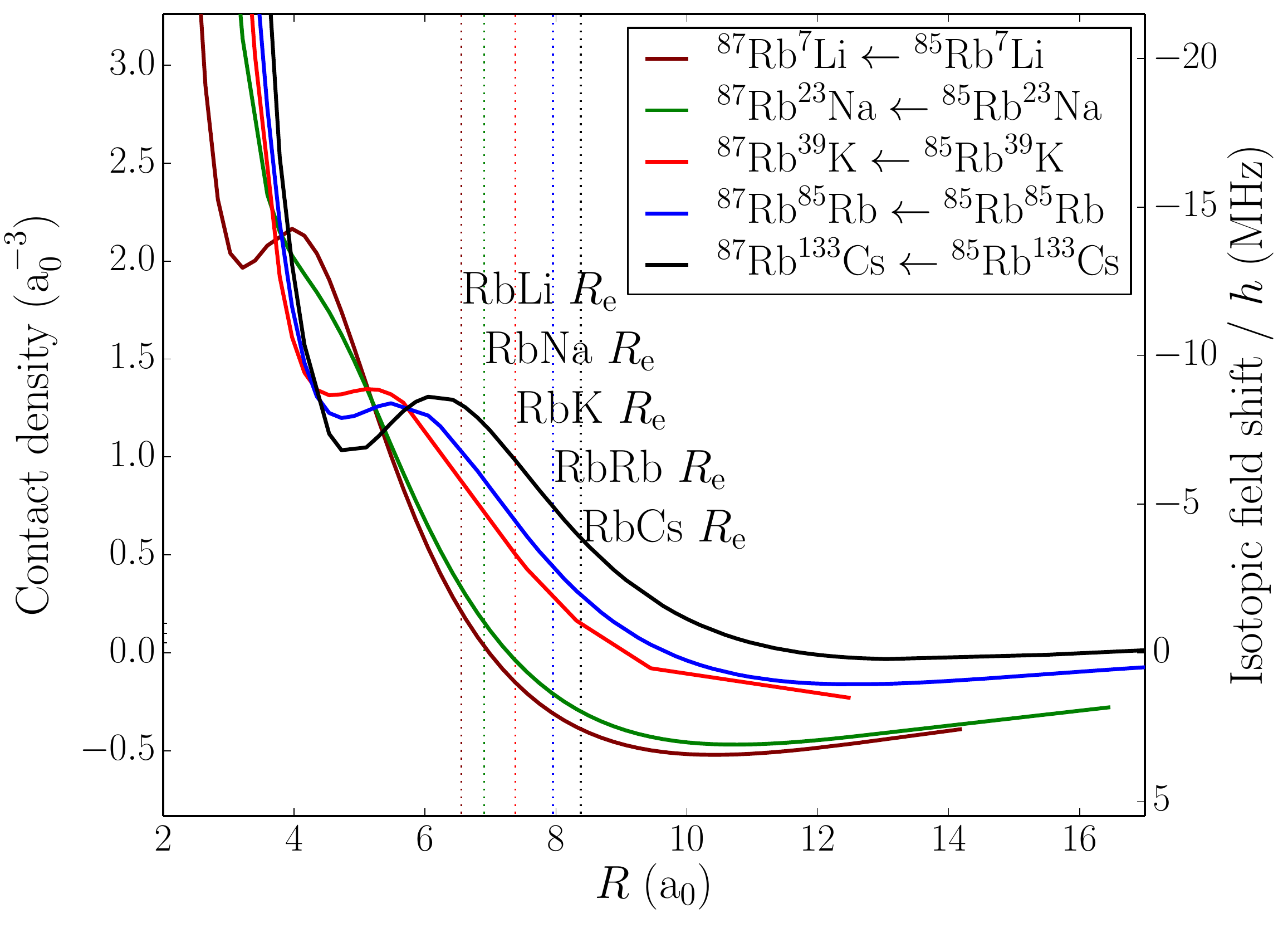} \label{fig:RbXFS} \\ (c)
\caption{Bonding-induced changes in contact densities and isotopic field shifts
in alkali-metal dimers. (a) Li nuclei; (b) K nuclei; (c) Rb nuclei. \label{fig:FSs}}
\end{figure}

The radial functions for bonding-induced changes in contact densities are shown
in Fig.\ \ref{fig:FSs} for the alkali-metal dimers. The field shifts for
$^6$Li$\leftarrow$$^7$Li, $^{41}$K$\leftarrow$$^{39}$K and
$^{87}$Rb$\leftarrow$$^{85}$Rb are shown explicitly on the right-hand axes.
Note that the changes in the nuclear charge radius are positive for
$^6$Li$\leftarrow$$^7$Li ($\delta\langle r^2 \rangle=0.731$~fm$^2$) and
$^{41}$K$\leftarrow$$^{39}$K ($\delta\langle r^2 \rangle=0.117$ fm$^2$), but
negative for $^{87}$Rb$\leftarrow$$^{85}$Rb ($\delta\langle r^2
\rangle=-0.0362$ fm$^2$).

The overall magnitude of the bonding-induced changes in contact densities
increases substantially from Li to K to Rb. This arises mostly because the
non-relativistic contact density for an $n$s electron is given approximately by
the Goudsmit-Fermi-Segr{\'e} (GFS) approximation \cite{Ko58,Ar71},
\begin{equation}
|\Psi(0)|^2 = \pi a_0^{-3} Z Z_a^2 \left(1 - \frac{d\sigma}{dn}\right) / n^{*3}.
\end{equation}
where $Z_a$ is the outer charge (which is 1 for Group I atoms), $n^{*} =
n-\sigma$ is the effective quantum number and $\sigma$ is the quantum defect.
The effective quantum number for the valence s electron increases only slowly
down Group I, taking values of 1.588, 1.626, 1.771 and 1.805 for Li, Na, K and
Rb, respectively \cite{vdSMbook}, while $d\sigma/dn$ is small. This gives an
overall contact density approximately proportional to $Z$. When electrons are
treated relativistically, heavy atoms accumulate an additional $Z$ scaling when
the factor $\alpha Z$ becomes non-negligible \cite{DFbook,Moss,BDNps2013}.

The overall shapes of the curves in Figs.\ \ref{fig:FSs}(a) to \ref{fig:FSs}(c)
can be explained in terms of simple bonding ideas. For the homonuclear
alkali-metal dimers, the contact density changes are negative at larger
distances, but positive at short range. This is qualitatively the same as the
behavior for H$_2$ \cite{Bader:1968}. The negative values at longer range
occurs partly because covalent bonding results in sp hybridisation, reducing
the population in the $n$s orbital and thus reducing the contact density. This
effect is partly counteracted by contraction of the core in response to the
reduced screening of the nucleus by p electrons. At short range, by contrast,
the two atoms interpenetrate one another sufficiently to increase the density
at both nuclei. For all the alkali-metal dimers, the overall contact density
change is positive at the equilibrium distance, but with a substantial negative
gradient.

For the heteronuclear dimers, there are additional effects from charge
transfer, which reduce the density on the electropositive atom and increase the
density on the electronegative one. These charge transfer effects counteract
the covalent reduction slightly for Li in LiNa and overwhelm it for Li in the
very polar molecules LiK, LiRb and LiCs. They also reinforce the covalent
reduction slightly for Rb in KRb and substantially for K and Rb in LiK, LiRb,
NaK and NaRb, and counteract it for K in KRb and KCs and for Rb in RbCs. All
these charge transfer effects correlate reasonably well with the corresponding
dipole moment functions \cite{Aymar:2005}.

\begin{figure}
\centering
\includegraphics[width=0.60\textwidth]{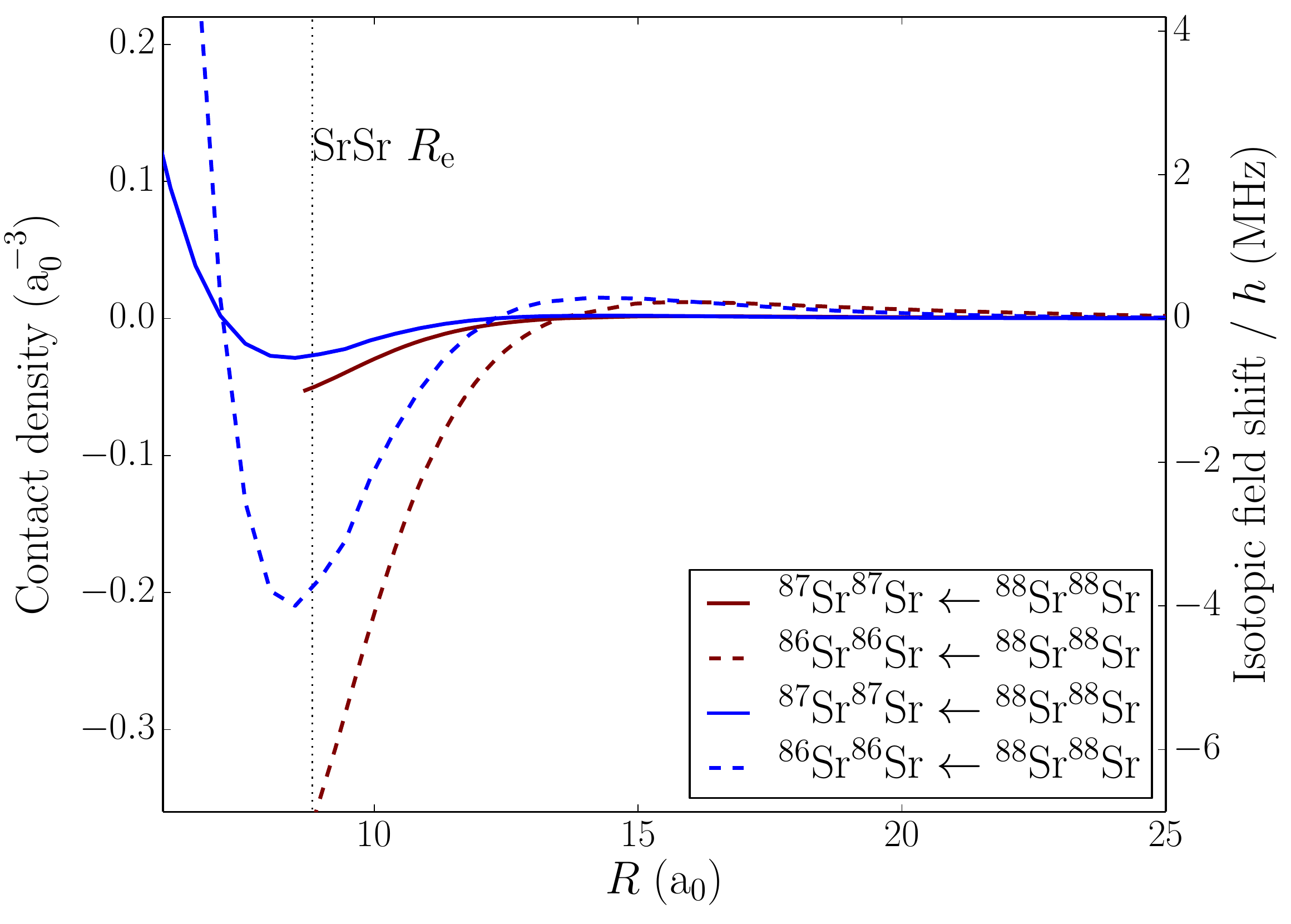} \\ (a) \\
\includegraphics[width=0.60\textwidth]{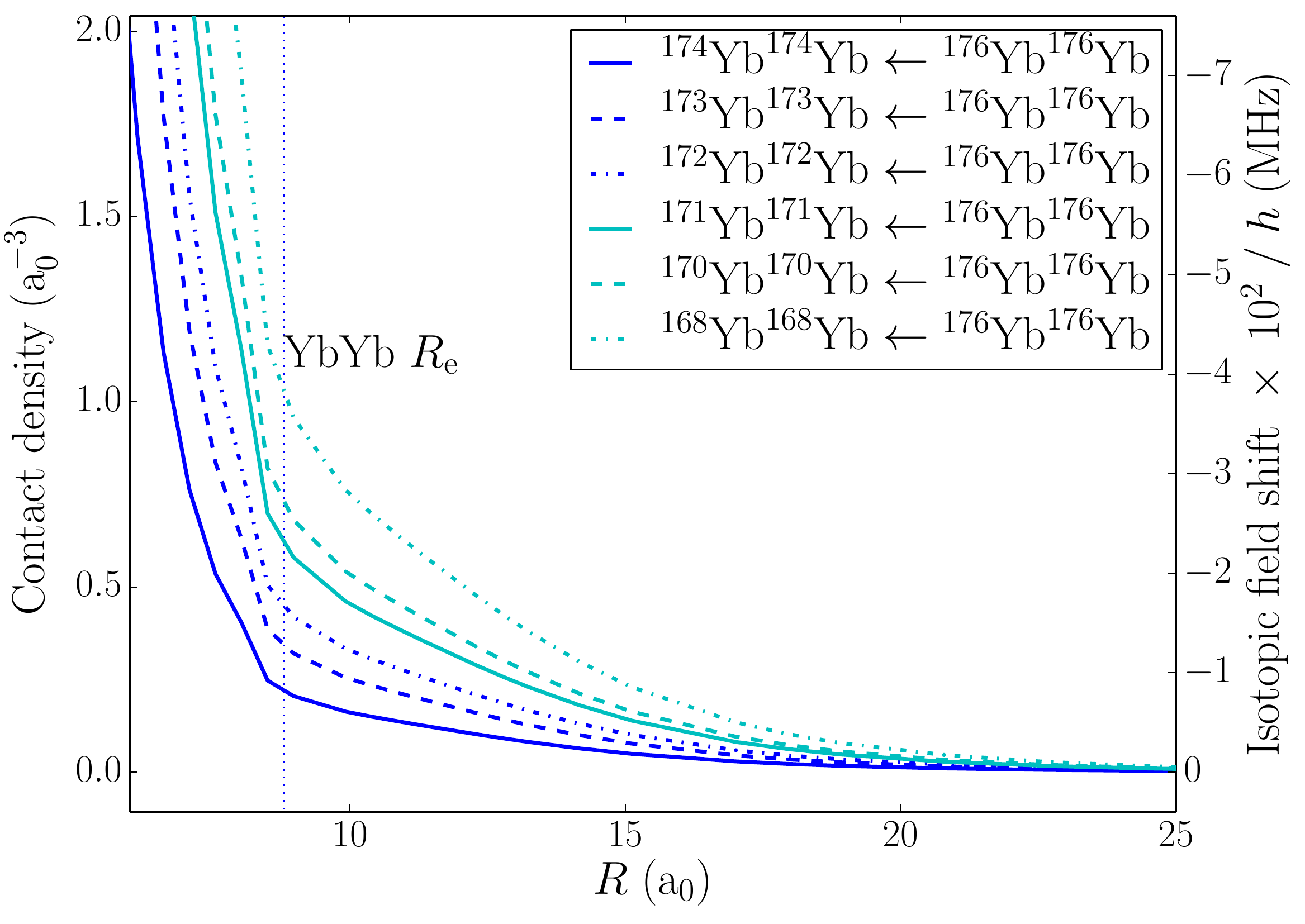} \\ (b)
\caption{Bonding-induced changes in contact densities and isotopic field shifts
in (a) Sr$_2$, as computed with 4-component DFT/B3LYP/v3z (maroon) and with
ZORA/B3LYP/QZ4P (blue); (b) Yb$_2$, with ZORA/B3LYP/QZ4P. The contact density
axis corresponds to the field-shift function for the largest isotopic
transition in each frame. \label{fig:SrYbFSs}}
\end{figure}

The situation is different for Sr$_2$, shown in Fig.\ \ref{fig:SrYbFSs}(a).
Here the contact density change due to bonding is positive at long range, but
becomes negative around $R_{\mathrm{e}}$, where there is weak chemical bonding.
The 4-component DFT/B3LYP/v3z calculations with DIRAC failed to converge at
internuclear distances less than 8.7 bohr, so we supplemented them with
ZORA/B3LYP/QZ4P calculations using ADF, shown in blue in Fig.\
\ref{fig:SrYbFSs}(a). These show a weaker negative feature, probably   a depth
of only $\approx400$ cm$^{-1}$ compared to the experimental value of 1081
cm$^{-1}$, though also perhaps because of the incomplete treatment of
relativity in ADF. At short range, the ADF contact densities turn upwards, due
to interpenetration of the atomic densities. Similar short-range behavior has
been seen theoretically in He$_2$ \cite{Bader:1968}, Ne$_2$ and Ar$_2$
\cite{Sebastian:1979}.

For Yb$_2$, 4-component DFT/B3LYP/v3z calculations with DIRAC failed to
converge entirely, so we used ZORA/B3LYP/QZ4P calculations with ADF instead.
The resulting bonding-induced contact density changes are shown in Fig.\
\ref{fig:SrYbFSs}(b). They are positive across the whole range of $R$, without
a negative region near $R_{\mathrm{e}}$. This may be simply because Yb$_2$
shows weaker covalent bonding than Sr$_2$. It may be noted that Yb shows much
stronger relativistic effects than Sr, and the contact density change for the
lowest $^1{\rm S_0}\rightarrow^1{\rm P}_1$ transition from multiconfiguration
Dirac-Fock calculations \cite{Torbohm:1985} is a factor of 5.3 larger. This is
reflected in the overall magnitude of the bonding-induced changes to contact
densities, which are much larger for Yb$_2$ than for Sr$_2$.

The increase in the bonding contributions to field shifts in moving from Li to
Yb is important, particularly when contrasted with the decrease in the mass shifts
through the same series. The effect of bonding on Li field shifts is a few
hundred kHz, while the mass shifts for the same systems are on the order of GHz.
However, the effects of bonding on field shifts for K and Rb are a few MHz,
while the mass shifts are on the order of tens of MHz. Thus, molecules formed from
heavier alkali metals may have comparable bonding contributions to both field
shifts and mass shifts, and studies of isotope shifts for these systems should
consider both effects. The same comments apply to Sr$_2$. Continuing the trend,
the effects of bonding on field shifts for Yb$_2$ are computed to be at least
two orders of magnitude larger than the corresponding mass shifts. The bonding
changes in field shifts for this system are on the order of tens of MHz.

The functions shown in Figs.\ \ref{fig:DBOCs} and \ref{fig:FSs} indicate that
there is indeed a crossover point in atomic mass where the effect of bonding on
the field shift becomes larger than on the mass shift. This is analogous to the
crossover between field and mass shifts for atoms \cite{BDNps2013}, and should
be considered in studies that consider molecular isotope shifts beyond the
usual mass scaling.

\subsection{Comparison with empirical isotope shifts}
\label{sec:emp}

When characterizing a particular electronic state spectroscopically, it is
sometimes possible to isolate small mass-dependent effects not accounted for by
rudimentary mass-scaling. Such Born-Oppenheimer breakdown functions have been
derived empirically for a number of relatively light diatomic molecules by
least-squares fitting of measured line positions for a pair or series of
isotopologs \cite{Coxon:HeH+:1999, Coxon:BeH+:1997, LeRoy:1999, Coxon:HF:2006,
Coxon:HX:2015, Coxon:HCl:2000, LeRoy:2002, LeRoy:2005, Coxon:LiH:2004,
LeRoy:2006, Henderson:2013, Coxon:CO:2004, WMLjcp2002, Coxon:2006, LeRoy:2009,
Tiemann:2009, Ivanova:2011}. The purpose of this section is to compare examples
of Born-Oppenheimer breakdown functions from the literature with the results
obtained using {\em ab initio} techniques. In particular, we will describe
results for the Li$_2$, LiK, and LiRb systems, since these are the only
alkali-metal diatomics for which we could find Born-Oppenheimer breakdown
functions in the spectroscopic literature. Since in Section \ref{sec:r_ifs} the
effects of chemical bonding on isotopic field shifts were found to be on the
order of kHz for the substitution $^6$Li $\leftarrow$ $^7$Li, we include only
mass shifts in this section.

For the $A^{1}\Sigma_u^+$ and $X^1\Sigma_g^+$ states of Li$_2$, extensive sets
of high-quality line positions have been measured for the
$^7$Li$^7$Li, $^6$Li$^7$Li, and $^6$Li$^6$Li isotopologs, and have been used
in several studies to derive correction functions for Born-Oppenheimer
breakdown. Since $^7$Li is the most abundant isotope, it is used in the
spectroscopic literature as the reference isotope. The $^7$Li$^7$Li
isotopolog is the one for which the most line positions have been measured,
so has the best-determined potential curve. In the following, the correction
functions are given for the substitution $^6$Li$^7$Li$\leftarrow$$^7$Li$^7$Li.

\begin{figure}
\begin{center}
\includegraphics[width=.52\textwidth]{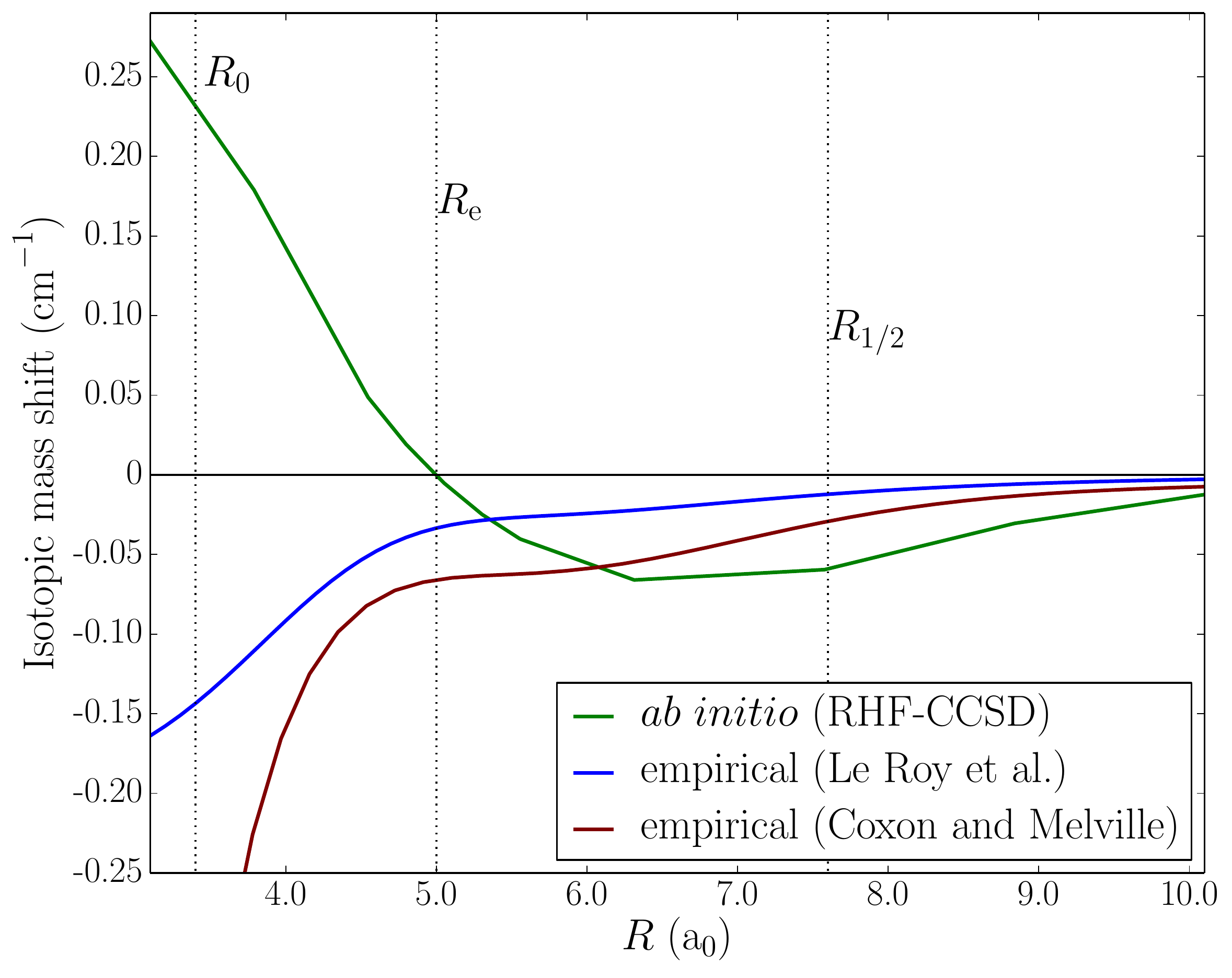}
\caption{Comparison of $^6$Li$^7$Li$\leftarrow$$^7$Li$^7$Li {\em ab initio}
mass-shift functions $\delta V^{\rm ad}(R)$ as computed at the RHF-CCSD/ANO-RCC level
of theory with empirical BOB functions obtained in Refs.\ \cite{Coxon:2006} and
\cite{LeRoy:2009}. \label{fig:LeRoy}}
\end{center}
\end{figure}

Figure \ref{fig:LeRoy} compares our mass-shift function $\delta V^{\rm
ad}_{\beta\leftarrow\alpha}(R)$, computed at the RHF-CCSD/ANO-RCC level of
theory, with the empirical Born-Oppenheimer breakdown corrections developed by
Coxon and Melville \cite{Coxon:2006} and by Le Roy {\em et al.}\
\cite{LeRoy:2009}. All three functions are negative in the long-range region,
with the function asymptotically approaching zero from below. However, the {\em
ab initio} function changes sign near the potential minimum and is positive at
short range, rising steeply between the potential minimum and the inner turning
point at the dissociation energy. The empirical functions, by contrast, remain
negative at short range.

In comparing the {\em ab initio} and empirical correction functions, it is
important to appreciate that the {\em ab initio} function represents a true
adiabatic correction, whereas the empirical functions in Fig.\ \ref{fig:LeRoy}
are {\em effective} adiabatic corrections that contain contributions from both
adiabatic and nonadiabatic terms. Watson \cite{Watson:1980} showed that it is
not possible to separate the adiabatic and nonadiabatic contributions on the
basis of line positions alone. However, an alternative formulation of the
Hamiltonian by Herman and Ogilvie \cite{Herman:1998} does allow the
contributions to be separated, using constraints from the molecular dipole
moment function or rotational $g$-factor. This approach has not yet been
applied to alkali-metal dimers, but Coxon and Hajigeorgiou have shown that, for
HCl \cite{Coxon:HCl:2000} and CO \cite{Coxon:CO:2004}, the empirical true and
effective adiabatic corrections have similar values near $R_{\rm e}$ but very
different gradients (actually of opposite sign for HCl). This may be the origin
of the qualitative differences in shape in Fig.\ \ref{fig:LeRoy}.

\begin{figure*}[t]
\begin{center}
\begin{minipage}{\textwidth}
\subfigure[]{
        \resizebox*{.52\textwidth}{!}{\includegraphics{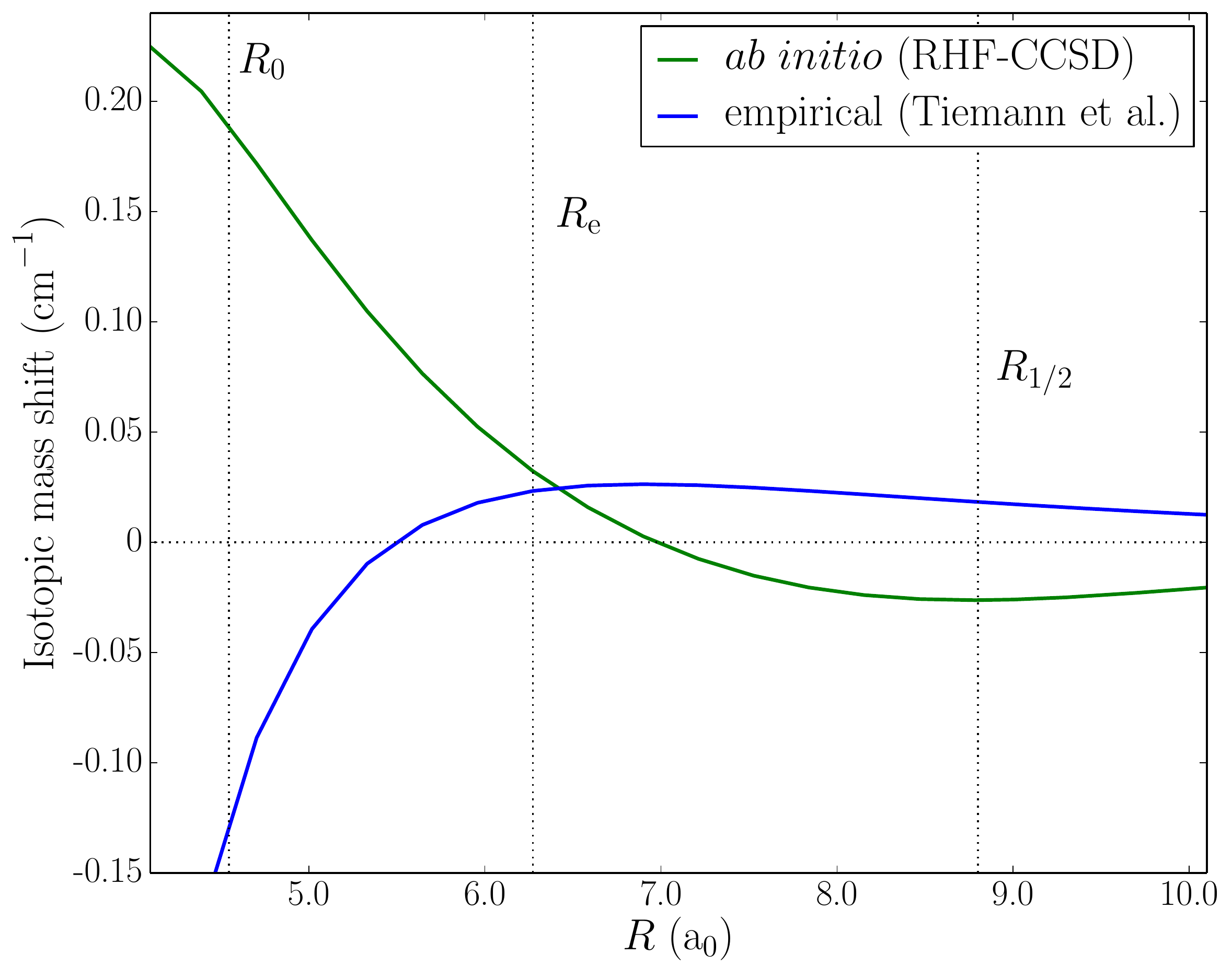}}
        \label{fig:Tiemann1}}
\subfigure[]{
        \resizebox*{.52\textwidth}{!}{\includegraphics{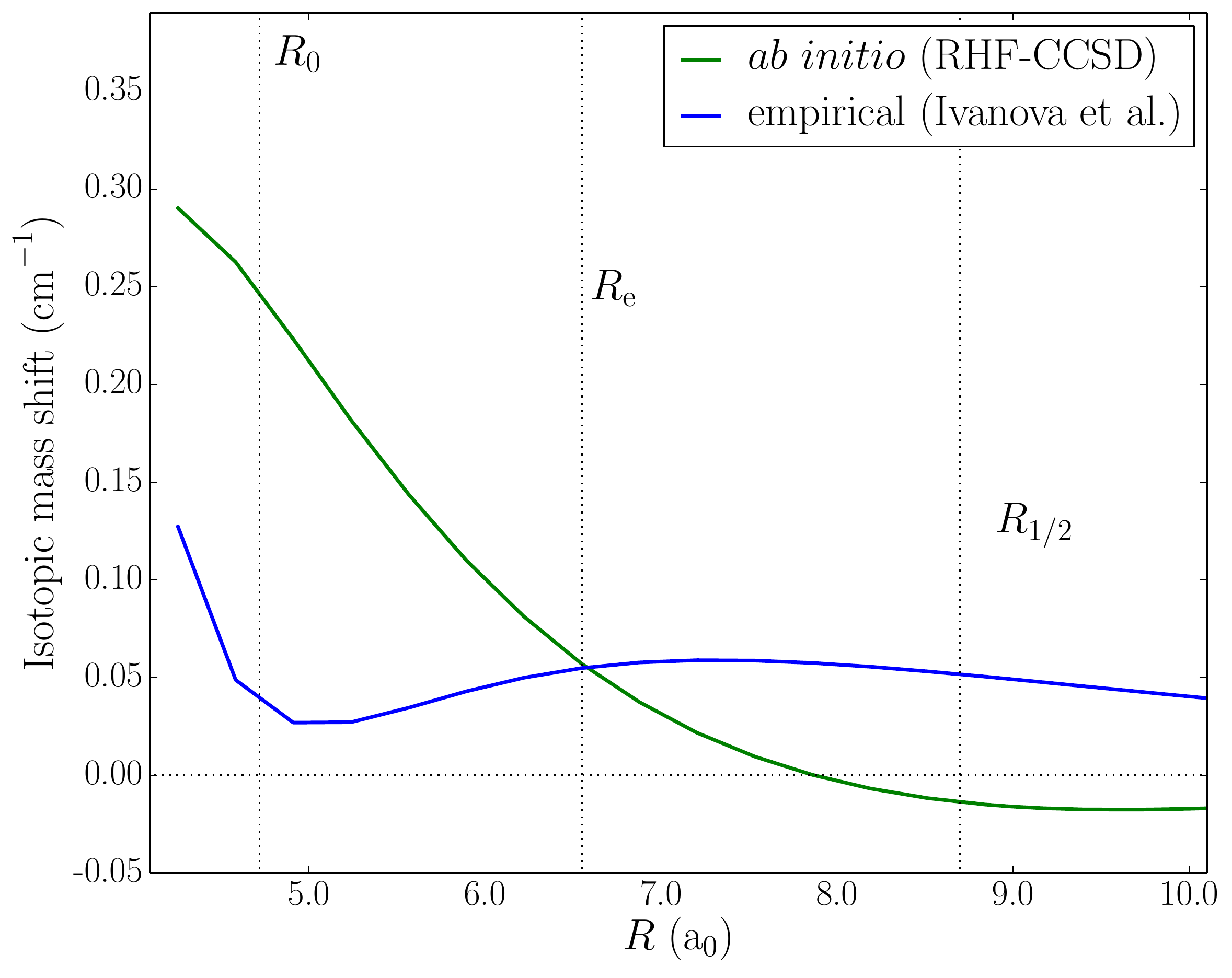}}
        \label{fig:Tiemann2}}
\caption{Comparison of {\em ab initio} $^{6}$Li$\leftarrow$$^{7}$Li mass-shift
functions $\delta V^{\rm ad}(R)$ as computed at the RHF-CCSD/ANO-RCC level of
theory with empirical BOB functions for (a) Li$^{39}$K \cite{Tiemann:2009} (b)
Li$^{85}$Rb \cite{Ivanova:2011}. \label{fig:Tiemann}}
\end{minipage}
\end{center}
\end{figure*}

We next consider LiK and LiRb, for which extensive spectra were measured in
Refs.\ \cite{Tiemann:2009} and \cite{Ivanova:2011} using high-resolution
fluorescence spectroscopy. The spectra were used to obtain ground-state
potential energy curves and Born-Oppenheimer breakdown functions by
least-squares techniques. Figure \ref{fig:Tiemann} compares the empirically
determined Born-Oppenheimer breakdown functions with our {\em ab initio} mass
shift functions, computed at the RHF-CCSD/ANO-RCC level of theory, with the
$^7$Li isotopolog taken as the reference species in each case. The {\em ab
initio} functions for LiK and LiRb have qualitatively similar features to that
for Li$_2$, which in turn is qualitatively similar to that for H$_2$. For both
systems, the {\em ab initio} and empirical functions have similar values at the
equilibrium distance $R_{\rm e}$, but different gradients and different overall
shapes. Tiemann {\em et al.}\ \cite{Tiemann:2009} comment that the adiabatic
correction near $R_{\rm e}$ for LiK is about 5 times larger than the typical
uncertainty in line positions at low $v$, so is statistically well determined.
As for Li$_2$, the difference in gradient between the {\em ab initio} and
empirical functions around $R_{\rm e}$ may plausibly be attributed to the fact
that the empirical functions represent effective adiabatic corrections that
include contributions from nonadiabatic effects.

At very long range, outside the Le Roy radius $R_{\rm LR}$ \cite{LeRoy:1973},
the potential $V(R)$ between atoms in S states dies off as $-C_6 R^{-6}$. The
dispersion coefficient $C_6$ is slightly different for different atomic
isotopes, so $\delta V^{\rm ad}(R)$ should also be proportional to $R^{-6}$
\cite{PJcp2012}. For alkali-metal atoms the difference between the $C_6$
coefficients is determined mostly by the valence s electron, whose wavefunction
dies off at long range as $\exp(-Z^*r/a_M)$. Here $Z^*$ is the effective
nuclear charge and $a_M=4\pi\epsilon_0\hbar^2/(\mu_M e^2)$ is an effective Bohr
radius for an electron with reduced mass $m_e M/(m_e+M)$, where $M$ is the
nuclear mass. An atom with a finite-mass nucleus has a larger $a_M$, larger
polarizability and larger (more attractive) $C_6$ coefficients that an atom
with an infinite-mass nucleus. This corresponds to $\delta V^{\rm ad}(R)$ being
negative at long range. The nonadiabatic contributions to the effective
adiabatic correction are proportional to $dV/dR$, so die off as $R^{-7}$ and
should not affect the long-range sign.

All the {\em ab initio} mass-shift functions are negative at very long range,
but the empirical functions for LiK \cite{Tiemann:2009} and LiRb
\cite{Ivanova:2011} are positive. In addition, $\delta V^{\rm ad}(R)/V(R)$
should approach a constant outside $R_{\rm LR}$. However, Fig.\ 4 of ref.\
\cite{Tiemann:2009} shows that the empirical $\delta V^{\rm ad}(R)/V(R)$ for
LiK increases nearly linearly between $R=15$ and 23 bohr, with little sign of
levelling off. It has reached only about 20\% of its asymptotic value at 23
bohr, which is well outside the Le Roy radius. We therefore conclude that the
long-range behaviour of the functional form used for the ground- and
excited-state Born-Oppenheimer breakdown (BOB) functions should be revisited in
future interpretations of the spectra.

\section{Isotope shifts in ultracold molecular physics}

The quantities that are usually measured in ultracold molecular physics are the
binding energies of levels very close to dissociation, often as a function of
magnetic field, and the positions of zero-energy Feshbach resonances as a
function of magnetic field. The levels of interest are often bound by only a
few MHz, which is less than 1 part in $10^7$ of the well depth for the
alkali-metal dimers. Because of this, they are very strongly dominated by
long-range effects, and are insensitive to the shape of the short-range
potential. Nevertheless, the binding energies of these levels depend
sensitively on the fractional part of the non-integer quantum number at
dissociation $v_{\rm D}$ \cite{LeRoy:1970}.

Considerable insight into the effects of small potential shifts may be gained
by writing the quantum number at dissociation semiclassically in terms of a WKB
phase integral,
\begin{equation}
\pi(v_{\rm D}+x)=\Phi,
\end{equation}
where
\begin{equation}
\Phi=\int_{R_0}^\infty \left(\frac{-2\mu V(R)}{\hbar^2}\right)^\frac{1}{2} dR,
\label{eq:phi}
\end{equation}
$V(R)$ is the internuclear potential for a reference isotopolog and $R_0$ is
the inner turning point at the dissociation energy. The usual WKB fraction
$x=1/2$ is replaced by $x=5/8$ at dissociation if $V(R)$ is asymptotically
$-C_6 R^{-6}$ \cite{Gribakin:1993}. The scattering length $a$ for a single
potential curve is related to the phase integral by
\begin{equation}
a=\bar{a}\left[1-\tan\left(\Phi-\frac{\pi}{8}\right)\right]
=\bar{a}\left\{1-\tan\left[\pi\left(v_{\rm D}+\frac{1}{2}\right)\right]\right\},
\end{equation}
where $\bar{a}$ is the mean scattering length of Gribakin and Flambaum
\cite{Gribakin:1993}, which depends only on $C_6$ and the reduced mass $\mu$.
The presence of $\mu$ in the integrand of Eq.\ \ref{eq:phi} produces the normal
Born-Oppenheimer mass scaling. If a mass-dependent perturbation $\delta V(R)\ll
V(R)$ is now introduced, $v_{\rm D}$ changes by an additional amount
\begin{equation}
\delta v_{\rm D}=\int_{R_0}^\infty
-\frac{1}{2\pi} \left(\frac{2\mu|V(R)|}{\hbar^2}\right)^\frac{1}{2}
\left(\frac{\delta V(R)-\delta V(\infty)}{|V(R)|}\right) dR, \label{eq:delta-vd}
\end{equation}
This integral is formally problematic because $V(R_0)=0$ so the condition
$\Delta V(R)\ll V(R)$ is not satisfied very close to the turning point for the
reference isotope, where $|V(R)|$ is comparable to $\delta V(R)$. In reality
this should be handled by a shift of $R_0$, but in practice this region makes
little contribution to the integral and can be neglected.

\begin{figure}
\begin{center}
\includegraphics[width=.52\textwidth]{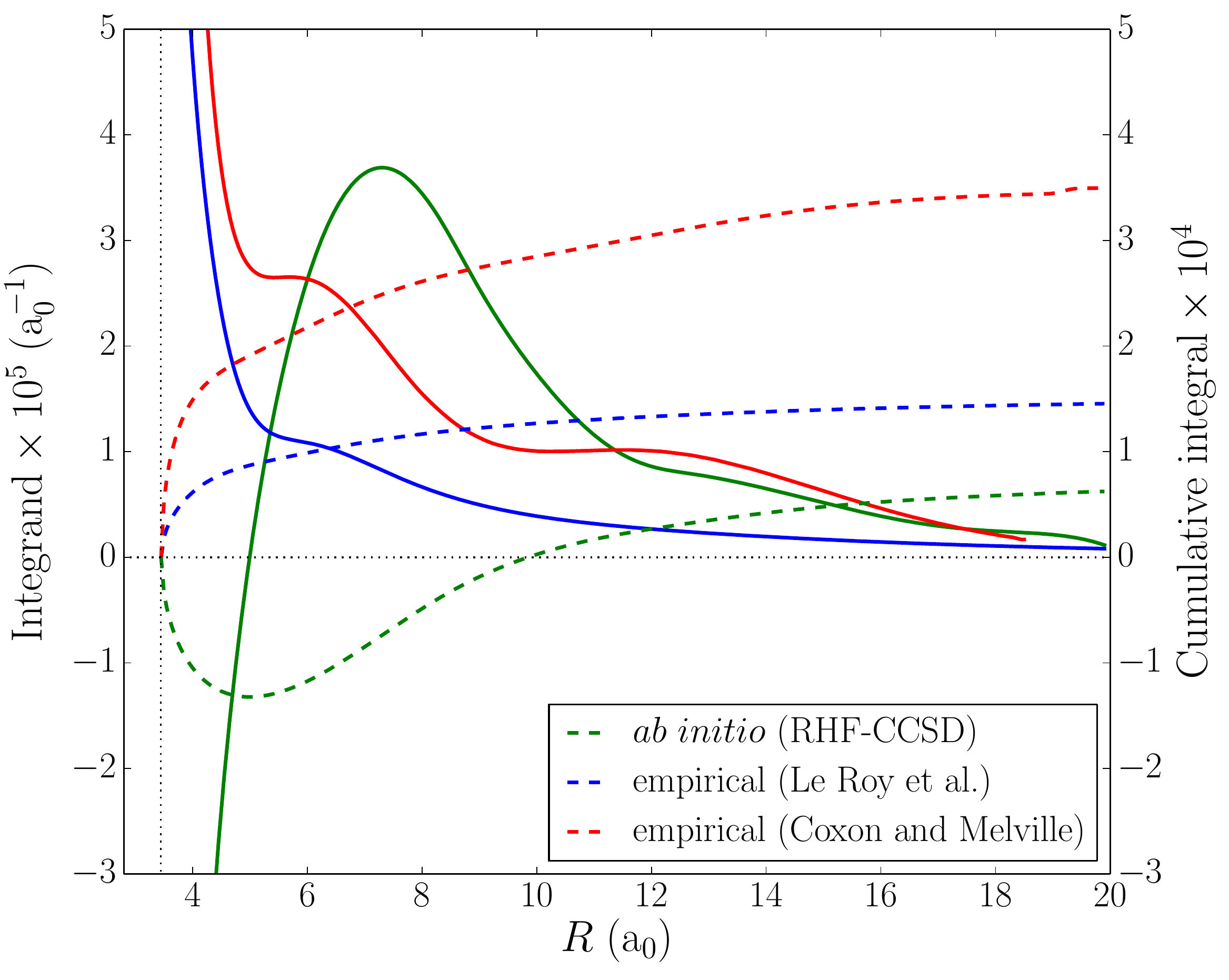}
\caption{Comparison of the semiclassical integrand of Eq.\ \ref{eq:delta-vd} for
$^6$Li$^7$Li$\leftarrow$$^7$Li$^7$Li (solid lines) and its cumulative integral
(dashed lines) obtained from {\em ab initio} theory (blue) with the semiempirical
functions of ref.\ \cite{Coxon:2006} (red) and ref.\ \cite{LeRoy:2009} (green).
\label{fig:LiLiintegrand}}
\end{center}
\end{figure}

Figure \ref{fig:LiLiintegrand} compares the integrand of Eq.\ \ref{eq:delta-vd}
for our mass-shift function $\delta V^{\rm ad}_{\beta\leftarrow\alpha}(R)$ for
$^6$Li$^7$Li$\leftarrow$$^7$Li$^7$Li, together with its cumulative integral,
with the corresponding quantities for the empirical functions of Coxon and
Melville \cite{Coxon:2006} and Le Roy {\em et al.}\ \cite{LeRoy:2009}. It may
be seen that all three give contributions to $\delta v_{\rm D}$ that are
positive for this substitution, but that the two empirical functions give
considerably larger values than the {\em ab initio} function. For comparison,
the analysis of Julienne and Hutson \cite{Julienne:Li67:2014} gave $\delta
v_{\rm D}=+9.4\times10^{-4}$ for $^6$Li$_2\leftarrow$$^7$Li$_2$ and would thus
give $\delta v_{\rm D}=+4.7\times10^{-4}$ for
$^6$Li$^7$Li$\leftarrow$$^7$Li$^7$Li. The {\em ab initio} function
underestimates this by a factor of about 5, but this might be because it
neglects nonadiabatic terms that contribute to the full effective adiabatic
correction. In addition, it should be noted that the value of the integral is a
delicate balance between short-range and long-range contributions.

\begin{figure*}[t]
\begin{center}
\begin{minipage}{\textwidth}
\subfigure[]{
        \resizebox*{.48\textwidth}{!}{\includegraphics{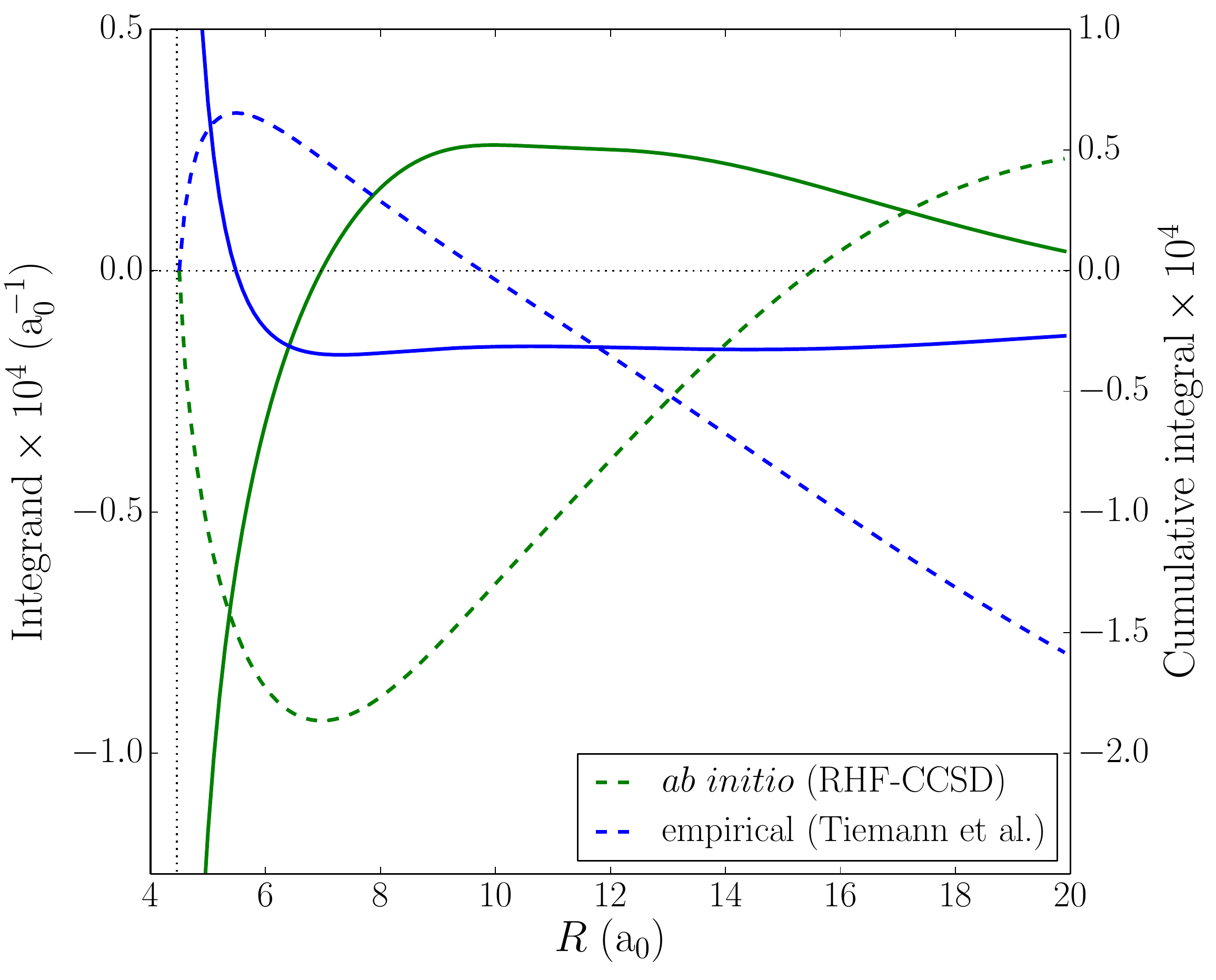}}
        \label{fig:LiKintegrand}}
\subfigure[]{
        \resizebox*{.48\textwidth}{!}{\includegraphics{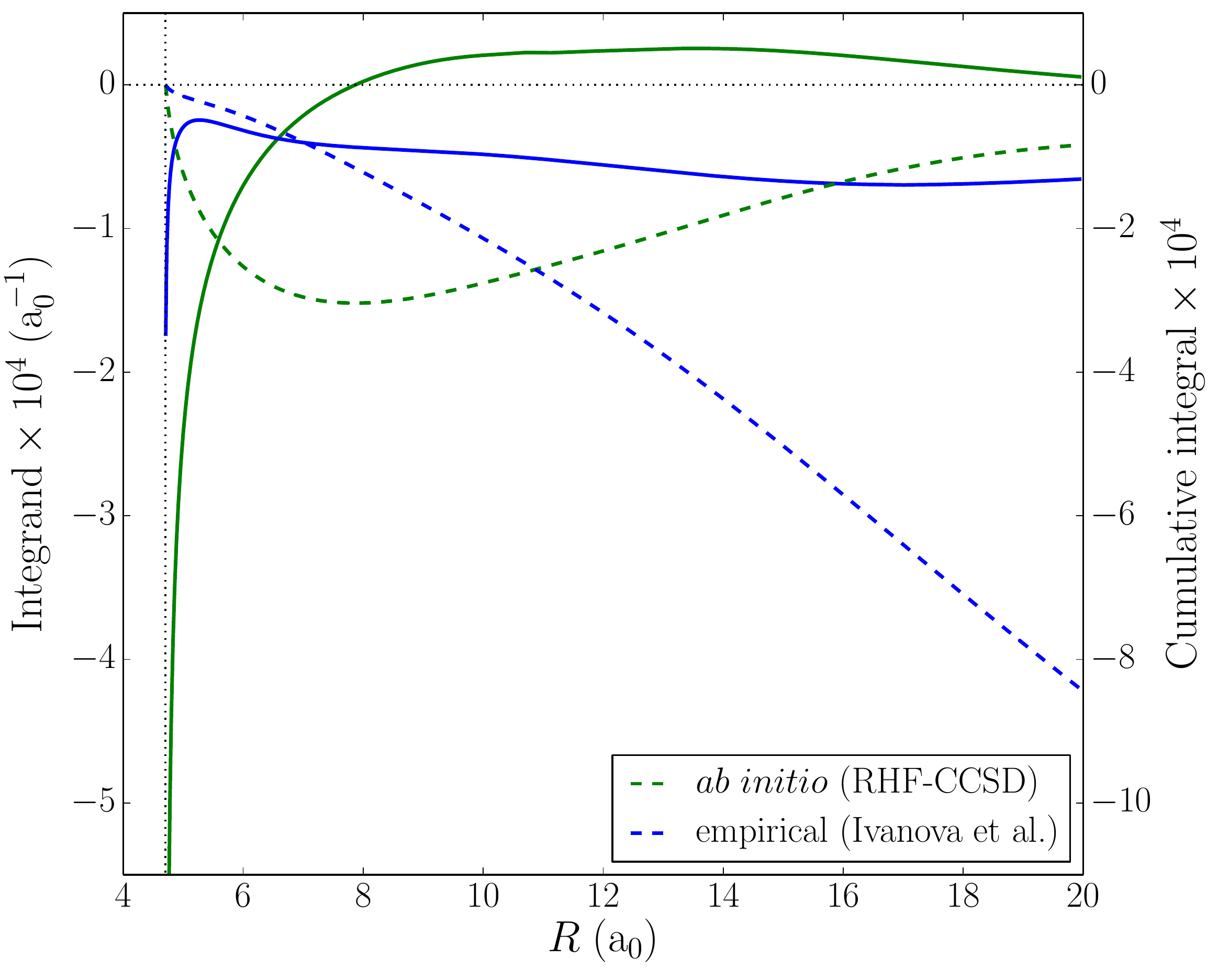}}
        \label{fig:LiRbintegrand}}
\caption{Comparison of the semiclassical integrand of Eq.\ \ref{eq:delta-vd}
(solid lines) and its cumulative integral (dashed lines) for
$^6$Li$^{39}$K$\leftarrow$$^7$Li$^{39}$K (panel a) and
$^6$Li$^{85}$Rb$\leftarrow$$^7$Li$^{85}$Rb (panel b) obtained from {\em ab
initio} theory (blue) with the semiempirical functions (green) of refs.\
\cite{Tiemann:2009} and \cite{Ivanova:2011}, respectively.
\label{fig:LiXintegrand}}
\end{minipage}
\end{center}
\end{figure*}

Figure \ref{fig:LiXintegrand} shows similar plots for the semiclassical
integrands and cumulative integrals in LiK and LiRb, comparing the {\em ab
initio} mass-shift functions with the empirical ones of refs.\
\cite{Tiemann:2009} and \cite{Ivanova:2011}. Outside the Le Roy radius, $\delta
V^{\rm ad}(R)/V(R)$ should approach a constant, as described above, so that the
semiclassical integrand should die off as $k \propto R^{-3}$. The integrands
for the {\em ab initio} functions do show this behaviour, but those for the
empirical functions remain nearly constant, so that the corresponding phase
integrals show no sign of converging by $R=20$ bohr. This is because $\delta
V^{\rm ad}(R)/V(R)$ for the empirical functions increases nearly linearly in
this region, and does not approach its asymptotic value until much larger
distances.

It would be very valuable to obtain spectra of near-threshold levels or
Feshbach resonance positions for LiK and LiRb for both $^6$Li and $^7$Li, in
order to find values of $\delta v_{\rm D}$ that can be used, with the
constraints on the long-range functions established here, to determine improved
adiabatic correction functions.

\section{Conclusions}
\label{sec:conc}

We have investigated electronic structure calculations of bonding contributions
to breakdown of the Born-Oppenheimer approximation for a range of molecules
important in ultracold physics. These include the homonuclear and heteronuclear
alkali-metal dimers and the Sr$_2$ and Yb$_2$ molecules. We have considered both
isotopic mass shifts (also known as diagonal Born-Oppenheimer corrections, DBOCs,
or adiabatic corrections) and isotopic field shifts (nuclear volume effects).

In a first step, we explored the performance of different electronic structure
methods for the elementary systems H$_2$ and Li$_2$. For these systems, it is
well known that potential curves from restricted Hartree-Fock (RHF)
calculations have incorrect dissociation behavior, whereas those based on
unrestricted Hartree-Fock (UHF) reference calculations can have the correct
behavior. However, for mass shifts we demonstrated inherent problems with
methods based on UHF references. We suggest that, when computing mass shifts
for systems that dissociate to open-shell monomers, unrestricted methods should
be avoided whenever possible. When there is no alternative to unrestricted
methods, care must be taken to avoid the occurrence of orbital instability
envelopes.

For all our target molecules, we computed isotopic mass shifts at the CCSD
level and field shifts from electron densities at the nucleus (contact
densities) obtained with relativistic DFT methods. For many of the alkali-metal
dimers, the mass-shift functions change sign in the vicinity of the equilibrium
distance, as a result of competing physical effects dominating in the short-
and long-range regions. It is thus difficult to identify periodic trends in the
magnitude of the values near equilibrium. For the contact densities, the
bonding changes are consistently positive at short range, but at longer range
result from a combination of covalent effects (which are always negative) and
charge-transfer effects, which are positive for the more electronegative atom
in a molecule but negative for its partner.

The magnitudes of the mass shifts decrease with increasing atomic number, while
the opposite is true for the field shifts. For Li, Na and K, mass shifts are
strongly dominant. For Rb and Sr, mass shifts are still generally larger than
field shifts, but the latter are not insignificant. For Yb, however, the field
shifts are dominant.

For the light molecules Li$_2$, LiK, and LiRb, we have compared our {\em ab
initio} mass-shift functions with Born-Oppenheimer breakdown functions fitted
to electronic spectra.  For LiK and LiRb the {\em ab initio} functions have
similar values to the empirical functions at the equilibrium distance, where
the empirical functions are most reliably determined. However, in all three
cases the {\em ab initio} functions have qualitatively different shapes from
the empirical functions away from equilibrium. This may be because the
empirical functions are {\em effective} adiabatic corrections that include
contributions from nonadiabatic terms.

The {\em ab initio} functions are slightly more attractive at long range for
$^6$Li than for $^7$Li, as expected from the larger polarizability of $^6$Li.
The empirical functions for LiK and LiRb have the opposite sign to the {\em ab
initio} functions at long range. The results presented here should help inform
the qualitative shape of the functional form used in future analyses of
electronic spectra to model Born-Oppenheimer breakdown functions.

We also considered the effect of Born-Oppenheimer corrections on quantities of
interest in ultracold physics. Scattering lengths and the positions of
near-threshold levels may be related to the non-integer quantum number at
dissociation. We developed a theory based on semiclassical phase integrals to
give insight into how small perturbations affect this quantity. For Li$_2$, LiK
and LiRb, the overall effect arises from a subtle balance of short-range and
long-range effects. Neither the {\em ab initio} function nor the empirical
functions for Li$_2$ are in quantitative agreement with the overall mass shift
obtained from studies of Feshbach resonances and near-threshold bound states. A
simultaneous treatment of both types of experiment, incorporating insights from
the {\em ab initio} studies, is needed to resolve the remaining discrepancies.

For molecules such as Sr$_2$ and Yb$_2$, there are proposals to use deviations
from Born-Oppenheimer mass scaling to probe a possible ``fifth force" that may
exist in addition to the familiar electromagnetic, gravitational and strong and
weak nuclear forces. One possible force is a ``short-range gravity",
proportional to the product of the nuclear masses in the molecule. However,
before attributing any deviations from Born-Oppenheimer mass scaling to such
forces, it is crucial to consider effects due to conventional isotopic mass
shifts and field shifts. These have mass dependences different from short-range
gravity but are likely to be difficult to distinguish in experiments.
Nevertheless, if their effects can be calculated reliably, they can be taken
into account, providing greater sensitivity to a fifth force (or allowing a
tighter bound to be placed upon it). This work has provided an initial attempt
to investigate the magnitude of mass and field shifts in Sr$_2$ and Yb$_2$.

Data underlying this article are available at
http://dx.doi.org/10.15128/r2vd66vz89.

\section*{Acknowledgments}

This paper is dedicated to Robert J. Le Roy, who was JMH's postdoctoral
supervisor in 1981-83 and had a formative influence on his early scientific
career. Anyone who has had a draft paper ``Le Royed" will appreciate its
importance.

The authors are grateful for the use of the EPSRC UK National Service for
Computational Chemistry Software (NSCCS) at Imperial College London, awarded as
NSCCS Project CHEM700. We thank Paul Julienne for initiating this work and John
Coxon for bringing the importance of nonadiabatic effects to our attention. We
are also grateful to John Stanton, P\'eter Szalay, and Markus Reiher for
helpful discussions regarding specific functionalities, keywords, and
limitations associated with their software. This work was supported by the
Engineering and Physical Sciences Research Council [grant number EP/I012044/1].

\bibliographystyle{tMPH}
\bibliography{../all,jl-bib}
\end{document}